\documentclass{iopart}

\usepackage{booktabs}
\usepackage{graphicx}
\usepackage{tabularx}
\usepackage{mathtools,amssymb,amsfonts}
\usepackage{comment}
\usepackage{float}
\usepackage{bm}
\usepackage{physics}
\usepackage{color}
\usepackage{placeins}
\usepackage{multirow}
\usepackage{booktabs}
\usepackage{siunitx}
\usepackage{tikz}
\usepackage{makecell}
\usepackage[svgnames]{xcolor}
\usepackage[colorlinks,
	linkcolor=Orange,
	citecolor=SlateBlue,
	urlcolor=DodgerBlue]{hyperref}
\usepackage{cleveref}
\graphicspath{{Figures/}}
\usepackage{orcidlink}
\usepackage{titlesec}

\newcommand{\appref}[1]{Appendix~\ref{#1}}

\begin{document}

\title{Fast Control for LIGO Interferometer Lock Acquisition}

\author{Masaya~Ono$^{1,2,\star}$\orcidlink{0009-0009-2138-2427}, Nathan A. Holland$^{1}$\orcidlink{0000-0003-1241-1264}, Radhika Bhatt$^{1}$\orcidlink{0009-0005-8354-906X}, Peter G. Carney$^{1}$\orcidlink{0000-0001-9351-9028}, Francisco Salces C\'arcoba$^{1}$\footnote[1]{Present address: Department of Physics and Astronomy, University of New Mexico, Albuquerque, NM 87131, USA}
\orcidlink{0000-0001-7049-4438}, Koji Arai$^{1}$\footnote[2]{Present address: Blue Laser Fusion, Inc., Goleta, CA 93117, USA}\orcidlink{0000-0001-8916-8915}, Rana X. Adhikari$^{1}$\orcidlink{0000-0002-5731-5076}}

\address{$^1$ LIGO Laboratory, California Institute of Technology, Pasadena, CA 91125, USA}
\address{$^2$ Department of Physics, The University of Tokyo, Bunkyo, Tokyo 113-0033, Japan}

\ead{$^{\star}$masayao@caltech.edu}
\vspace{10pt}
\begin{indented}
\item[]\today
\end{indented}

\begin{abstract}
Gravitational wave observatories are large scale laser interferometers which are held on resonance to facilitate the detection of gravitational waves. Achieving this resonant condition is a time consuming process, and in the LIGO gravitational-wave observatories this procedure has a relatively low success rate. This limitation leads to a loss of observation time thereby reducing the number of observable astrophysical events. The primary cause of this problem is the coupling among the longitudinal degrees of freedom of the interferometer, particularly the coupling from the arm cavity lengths to the central degrees of freedom, such as the recycling cavities and the Michelson interferometer.
We demonstrate a simpler and faster lock acquisition scheme at the 40-meter prototype interferometer, located at the California Institute of Technology. This scheme is achieved by controlling the arm cavity lengths with high bandwidth during the lock acquisition process. In this paper, we describe how this fast lock acquisition is realized, compare it with the current LIGO lock acquisition method, and discuss how our framework can be implemented in LIGO.

\end{abstract}

\section{Introduction}
\label{sec:intro}

A gravitational wave is a phenomenon in the universe in which distortions in space-time propagate as waves, and was detected for the first time by LIGO~\cite{aasi2015advanced} in 2015. Currently, ground-based gravitational wave detectors, LIGO, Virgo~\cite{acernese2014advanced}, and KAGRA~\cite{somiya2012detector,aso2013interferometer} are in operation targeting gravitational waves of which frequencies are $10 \sim 10^4 \, \rm{Hz}$. 
Future detectors such as Cosmic Explorer~\cite{abbott2017exploring} and Einstein Telescope~\cite{punturo2010einstein} are planned to be operational in the coming decades, and would likely require control systems similar to the existing ground-based interferometers.

These gravitational-wave observatories are, in essence, laser interferometers composed of suspended mirrors which form coupled optical cavities. 
To detect gravitational waves, it is necessary to precisely lock the interferometers on resonance by controlling the distances between the mirrors.
The process of locking all degrees of freedom of a gravitational wave interferometer into this resonant state, starting from a free-running state, is denoted lock acquisition~\cite{staley2014achieving, acernese2004lock,acernese2008lock,Akutsu_2020}.
The laser interferometer in LIGO is a Dual-Recycling Fabry-P\'{e}rot Michelson interferometer (DRFPMI) with five longitudinal control degrees of freedom~\cite{izumi2016advanced}: the common and differential arm cavity modes (CARM and DARM), the length degrees of freedom of the power and signal recycling cavities (PRCL~\cite{masaki1998phd} and SRCL~\cite{meers1989frequency, mizuno1993resonant, heinzel1996experimental, heinzel1999advanced}), and the Michelson interferometer degree of freedom (MICH). 
These degrees of freedom are controlled using RF signals derived from the phase modulation of the laser light. 
Variants of the widely used Pound-Drever-Hall (PDH)~\cite{drever1983laser} are used for reading out the multiple coupled degrees of freedom~\cite{izumi2016advanced}. The strong coupling between degrees of freedom and the narrow linear range of the signals make it extremely challenging to achieve optical resonances of the full system~\cite{Eric2018phd, Robert2010phd, Miyakawa_2006}.

To overcome this, LIGO uses an Arm Length Stabilization system (ALS)~\cite{kiwamu2012phd, izumi2012multicolour, mullavey2012als} that allows the resonance to be achieved in a deterministic way, for 2 of the degrees of freedom.
In the LIGO scheme, the 3 degrees of freedom in the compound Michelson achieve resonance in a stochastic way and then the 4\,km arms are brought to resonance slowly. This latter step is complicated, slow, and somewhat fragile which leads to a loss in duty cycle as detailed in \Cref{sec:LIGOLockAcq}. When the CARM offset reduction fails partway through, the interferometer must return to its down state, relock the input mode cleaner and the central interferometer, and reattempt from the beginning---a cycle that can repeat several times and substantially extends the total lock acquisition time.

The key concept of our fast lock acquisition scheme is to expand the bandwidth of the ALS CARM control loop. 
This suppresses the contamination from CARM fluctuations into the PRCL and MICH, which allows us to acquire lock of the power recycling Michelson interferometer with a very small CARM offset. 
Furthermore, unlike the current LIGO procedure, which requires several intermediate steps, such as gain adjustments when reducing the offset to zero, our framework, implemented on the California Institute of Technology (CIT) 40-meter prototype interferometer, can skip these steps. Reducing the number of control transitions results in a significant reduction in lock acquisition time due to the increased robustness of the offset reduction process.

This paper is organized as follows. \Cref{sec:LIGOLockAcq} reviews the current lock acquisition scheme in Advanced LIGO (aLIGO). 
\Cref{sec:FastLockAcq} describes the lock acquisition sequence realized in the CIT 40-meter interferometer. 
\Cref{sec:fastLIGO} discusses the evaluation of the locked interferometer and how our scheme can be implemented in LIGO. 
\Cref{sec:Conclusion} summarizes this work and outlines future prospects.

\section{Advanced LIGO Lock Acquisition}
\label{sec:LIGOLockAcq}

To compare aLIGO and the CIT 40-meter prototype we choose to focus on the lock acquisition for LIGO's main interferometer, the DRFPMI. There are other subsystems, such as the squeezing filter cavity, which need to be locked in aLIGO but are not present at the CIT 40-meter. Consequently, with direct comparison being impossible discussion of these is omitted - lock acquisition for the squeezing filter cavity occurs in parallel to the DRFPMI locking \cite{capote2025aligoO4} and thus does not add significant time.

This treatise specifically demonstrates an alternative to one stage of the existing lock acquisition procedure. To provide a more complete comparison, we choose to emphasize the relevant component of this sequence, CARM offset reduction. We analyze the limitations of the existing technique to highlight the potential benefit of our proposal. Subsection \ref{subsec:LIGOLockSum} provides a summary of the current aLIGO lock acquisition procedure, and subsection \ref{subsec:LIGOCarmOff} focuses on time accounting for the stages of this sequence our proposal would supersede.

\subsection{Locking the Dual Recycled Fabry-P\'{e}rot Michelson Interferometer}
\label{subsec:LIGOLockSum}

Lock acquisition for the main, DRFPMI component of aLIGO is a sequence with four distinct stages \cite{staley2014achieving, capote2025aligoO4, martynov2015phd}. In the first stage, ALS, with wavelength of 532 nm, is used to lock the Fabry-P\'{e}rot arms. Control of these optical cavities is transitioned to the CARM and DARM basis, and a locking point is chosen where the 532 nm and 1064 nm light are co-resonant in these interferometers.

Next, a longitudinal offset is applied to the arms in the CARM basis. This offset, denoted the CARM offset, holds the arms away from the 1064 nm resonance point, and control using 532 nm interferometry suppresses the length fluctuations with a bandwidth of $\approx$ 100 Hz \cite{martynov2015phd}. Combined, these two effects create a stable and predictable optical system for proceeding with lock acquisition.

With this CARM offset applied, the central, Dual-Recycling Michelson Interferometer (DRMI) is locked consisting of the power and signal recycling cavities, and the central Michelson interferometer. These are locked first with a standard, 1f, sideband technique before being transitioned to a 3f demodulation technique \cite{ARAI200015,arai2002tama300,arai2001robust}. The 3f lock is critical to ensure that the three central interferometric degrees of freedom remain resonant, providing reduced sensitivity to sign reversal of the arms' compound, composite reflectivity which flips while the previously applied CARM offset is zeroed.

Reducing the CARM offset is an involved process where several intermediate configurations are transitioned through before it is zeroed \cite{staley2014achieving,martynov2015phd}. This procedure is required by the residual motion the ALS lock provides, which exceeds the linewidth of the Power Recycled Common Arm cavity \cite{staley2014achieving}. During this process CARM control is transitioned from the ALS, to transmitted power, through a RF control signal normalized by transmitted power, and finally only to the RF control signal \cite{martynov2015phd}. Similarly, the DARM control is transitioned from ALS to RF and finally RF, normalized by transmitted power, error signals. At each stage, the residual motion present in the arm cavities is reduced as a more sensitive error signal is utilized for locking. When the CARM offset is zeroed, the central interferometric degrees of freedom are swapped from their 3f locking signals to 1f locking signals.

With all of the, DRFPMI, interferometric degrees of freedom locked, on first order (1f) sidebands, and the 1064 nm carrier resonant the final stages of lock acquisition commences - these have no analog in the CIT 40m prototype so are only briefly summarized. Adjustments are made to the input optical power, frequency shaping of the longitudinal and angular interferometric controls, and frequency-dependent squeezing is injected into the interferometer. To produce the low noise gravitational-wave readout the output mode cleaner is locked, and DARM is transitioned to DC readout \cite{Ward_2008, fricke2012dc}. Finally, the detectors are ready to observe gravitational waves.

\subsection{Time Accounting for CARM Offset Reduction in aLIGO}
\label{subsec:LIGOCarmOff}

As described in \Cref{subsec:LIGOLockSum}, LIGO's CARM offset reduction is a phased process, with both CARM and DARM control transitioned through various error signals. Since the second observing run this lock acquisition process is automated and managed by a state based algorithm called Guardian \cite{rollins2016guardian}. Each state in a Guardian node alters the control configuration of the interferometer and is provisioned with a unique state number.
Both LIGO Hanford Observatory (LHO) and LIGO Livingston Observatory (LLO) have a Guardian node that manages their interferometers overall state \cite{LHO_ISCgrd,LLO_ISCgrd}.

We use the mean, second trend of this Guardian state number to construct a time budget for CARM offset reduction at both LHO and LLO.
These trends are downloaded for observing runs O4a (May 24, 2023 to January 16, 2024), O4b (April 10, 2024 to January 28, 2025), and O4c (January 28, 2025 to November 18, 2025). Data segments that were unavailable, such as substantial portions of LHO data during O4c, were NaN-padded, and this padding was then discarded in our analysis.

Commencement and conclusion of CARM offset reduction is determined by thresholding and then detection of the rising and falling edges, respectively. Locklosses occurring during this procedure can cause the lock acquisition procedure to restart multiple times before CARM offset reduction successfully completes. This leads to several interpretations for the CARM offset reduction time. We choose two methods to pair CARM offset commencement and completion times, resulting in both extrema for the duration. Each procedure requires the commencement time to be later than the prior completion time and sooner than the completion time.

\begin{figure*}
    \includegraphics[width = 1.0\linewidth]{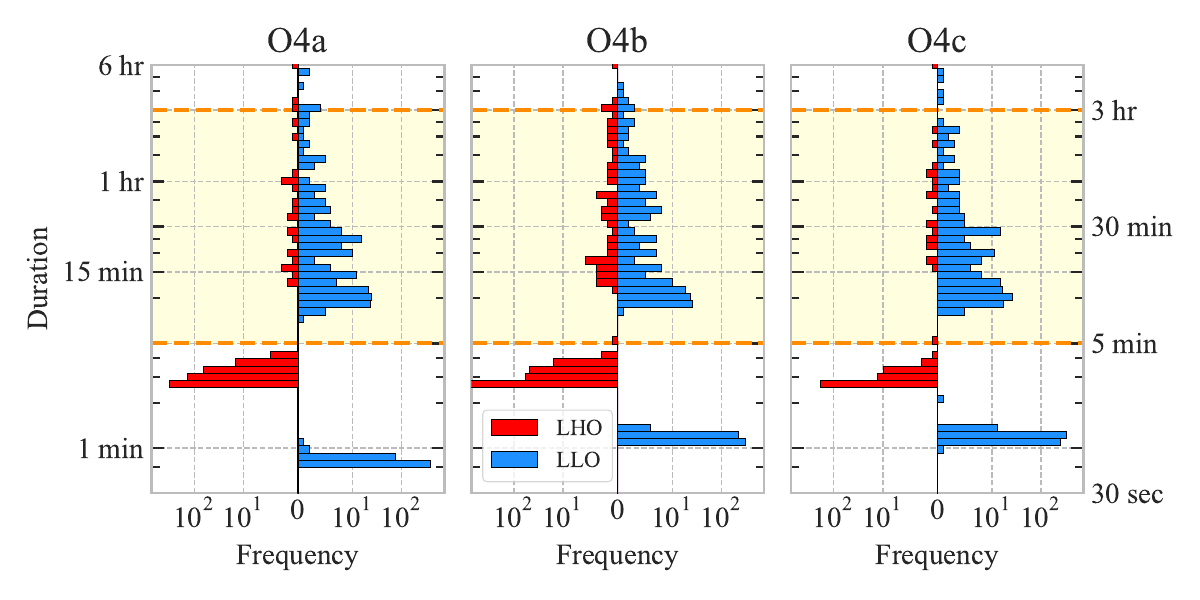}
    \caption{Histogram of \textit{with lockloss/es} CARM offset reduction times for LIGO during observing run four. CARM offset reduction commences when the second trend, the mean value of the observatories' ISC\_LOCK Guardian first exceeds a threshold value - 200 for LHO and 560 for LLO. The process ends when this value finally exceeds a second threshold value - 410 for LHO and 800 for LLO. If one or more locklosses occur during this process, the whole duration is attributed to CARM offset reduction. A primary distribution, around 3 minutes for LHO and 1 minute for LLO, corresponds to one-shot CARM offset reduction. The long tails, exceeding 5 minutes, are predominantly caused by one or more locklosses during CARM offset reduction.  The region less than 3 hours, represents situations where the technique presented in this paper will increase observation times.
    }
    \label{fig:LIGOlockAcq}
\end{figure*}

Our primary method associates all locklosses and downtime with CARM offset reduction, provided the first lockloss occurred during the CARM offset reduction process. This is denoted \textit{with lockloss(es)}, and the earliest commencement time is selected for this data set. Consequentially this represents the ceiling for CARM offset reduction durations. \Cref{fig:LIGOlockAcq} illustrates this distribution, and \Cref{tab:LIGOlockAcq} provides some of the key statistics.
Our alternative method does not permit lockloss to be associated with CARM offset reduction. It is denoted \textit{w/o lockloss(es)}, and the latest commencement time is selected for this data set. Correspondingly this represents the floor for CARM offset reduction durations. Key statistics are also summarized in \Cref{tab:LIGOlockAcq} and contrast with the \textit{with locklosses} data set.

\begin{table}
    \caption{Key statistics for CARM offset reduction at LIGO, for both sites during observing runs O4a, O4b and O4c. Median values exhibit little substantial change between the \textit{with lockloss/es} and \textit{w/o lockloss} data sets. Comparatively there is a substantial increase in standard deviation and skewness, except where the latter is already high. This indicates that the inclusion of lock losses, which occur during CARM offset reduction, substantially increases the duration attributed to this procedure. For durations less than three hours the observing time lost is estimated. Note that a significant fraction of data from LHO during O4c was unavailable and is excluded from analysis.}
    \centering
    \begin{tabular}{c|ccc|ccc}
        \toprule
        \multirow{2}{*}{\textbf{Statistic}} 
        & \multicolumn{3}{c|}{\textbf{\makecell{LIGO Hanford}}}
        & \multicolumn{3}{c}{\textbf{\makecell{LIGO Livingston}}} \\
        & O4a & O4b & O4c & O4a & O4b & O4c \\
        \midrule
        
        \textbf{Number} & 578 & 964 & 235 & 666 & 710 & 772 \\
        
        \midrule
        
        \makecell{\textbf{Median}\\$[$with lockloss(es)$]$}
         & 170 s & 160 s & 160 s & 50 s & 70 s & 70 s \\
        
        \makecell{\textbf{Standard Deviation}\\$[$with lockloss(es)$]$}
         & 5,400 s & 9,000 s & 1,700 s & 8,800 s & 4,500 s & 8,600 s \\
        
        \makecell{\textbf{Fisher-Pearson Skewness}\\$[$with lockloss(es)$]$}
         & 14 & 29 & 10 & 11 & 12 & 18 \\
        
        \midrule
        
        \makecell{\textbf{Median}\\$[$w/o lockloss(es)$]$}
         & 170 s & 160 s & 160 s & 50 s & 70 s & 70 s \\
        
        \makecell{\textbf{Standard Deviation}\\$[$w/o lockloss(es)$]$}
         & 18 s & 69 s & 19 s & 3 s & 4 s & 4 s \\
        
        \makecell{\textbf{Fisher-Pearson Skewness}\\$[$w/o lockloss(es)$]$}
         & 3 & 27 & 1 & -6 & -10 & 10 \\
        
        \midrule
        
        \makecell{\textbf{Time Lost}}
        & \makecell{1 day\\9 hr\\31 min}
        & \makecell{3 days\\1 hr\\33 min}
        & \makecell{$\geq$\\22 hr\\6 min}
        & \makecell{3 days\\19 hr\\5 min}
        & \makecell{4 days\\3 hr\\54 min}
        & \makecell{3 days\\14 hr\\27 min} \\
        \bottomrule
    \end{tabular}
    \label{tab:LIGOlockAcq}
\end{table}

\Cref{fig:LIGOlockAcq} shows a main grouping of CARM offset reduction times at both observatories. For LHO this is just under three minutes, and at LLO it is approximately one minute. Cross referencing with \Cref{tab:LIGOlockAcq}, particularly the \textit{w/o lockloss} statistics, demonstrates that this grouping corresponds to single-pass CARM offset reductions. The standard deviation of the \textit{w/o lockloss} data sets also indicates that single-pass CARM offset reductions are tightly grouped.

\Cref{fig:LIGOlockAcq} also shows substantial, long CARM offset reduction times, in the \textit{with locklosses} data set. This is caused by failure(s) at any of the multiple, sequential control transitions within the CARM offset reduction procedure, causing the interferometer to lose lock. When this occurs, the interferometer must execute its full down-state sequence, relocking commences, and the CARM offset reduction is reattempted from the beginning. Repeating the process outlined in \Cref{subsec:LIGOLockSum} adds substantial time attributable to the CARM offset procedure.

\Cref{tab:LIGOlockAcq} quantifies how the distribution of CARM offset reduction times changes substantially between the \textit{w/o lockloss} and \textit{with lockloss} data sets. The center of the distributions remains unchanged; however, both standard deviations and skewness increase by including locklosses. The simultaneous increase of standard deviation and skewness indicates the introduction of outliers at long durations. This demonstrates that when locklosses occur in LIGO's current CARM offset reduction technique substantial time is added to the observatories' lock acquisition procedure.

We calculate observing time lost to locklosses during CARM offset reduction by comparing the \textit{with locklosses} data sets from LHO and LLO, and results from the CIT 40-meter prototype given in \Cref{sec:FastLockAcq}.
First site, CARM offset reduction times longer than three hours are discarded - here we assume that other commissioning, investigation(s) or extended downtime is ongoing. From the remaining durations, 58 seconds is subtracted to represent the CARM offset time at the CIT 40-meter prototype. Where this difference is negative, these segments are ignored - this exclusion results in corrections shorter than one hour per observing run. The cumulative lost time is then calculated and presented in \Cref{tab:LIGOlockAcq}. Observing time at LHO can be increased by at least 1 day, per observing run, with the exception of O4c where a substantial fraction of the data was unavailable, indicated by a \textit{greater than or equal to} symbol. Larger increases in observing time are expected at LLO where several days can be gained per observing run.

LIGO Livingston Observatory data also emphasize that the robustness of the CARM offset reduction technique is critical. Here, the current single-shot CARM offset reduction times closely match our new technique, and are faster than those at LHO. However, there is a substantially more prominent tail compared to LHO, corresponding to CARM offset reduction restarting due to locklosses. Consequently, the observing time lost over a whole observing run is larger.

\section{Fast Lock Acquisition}
\label{sec:FastLockAcq}

\begin{figure}
    \includegraphics[width=1.0\linewidth]{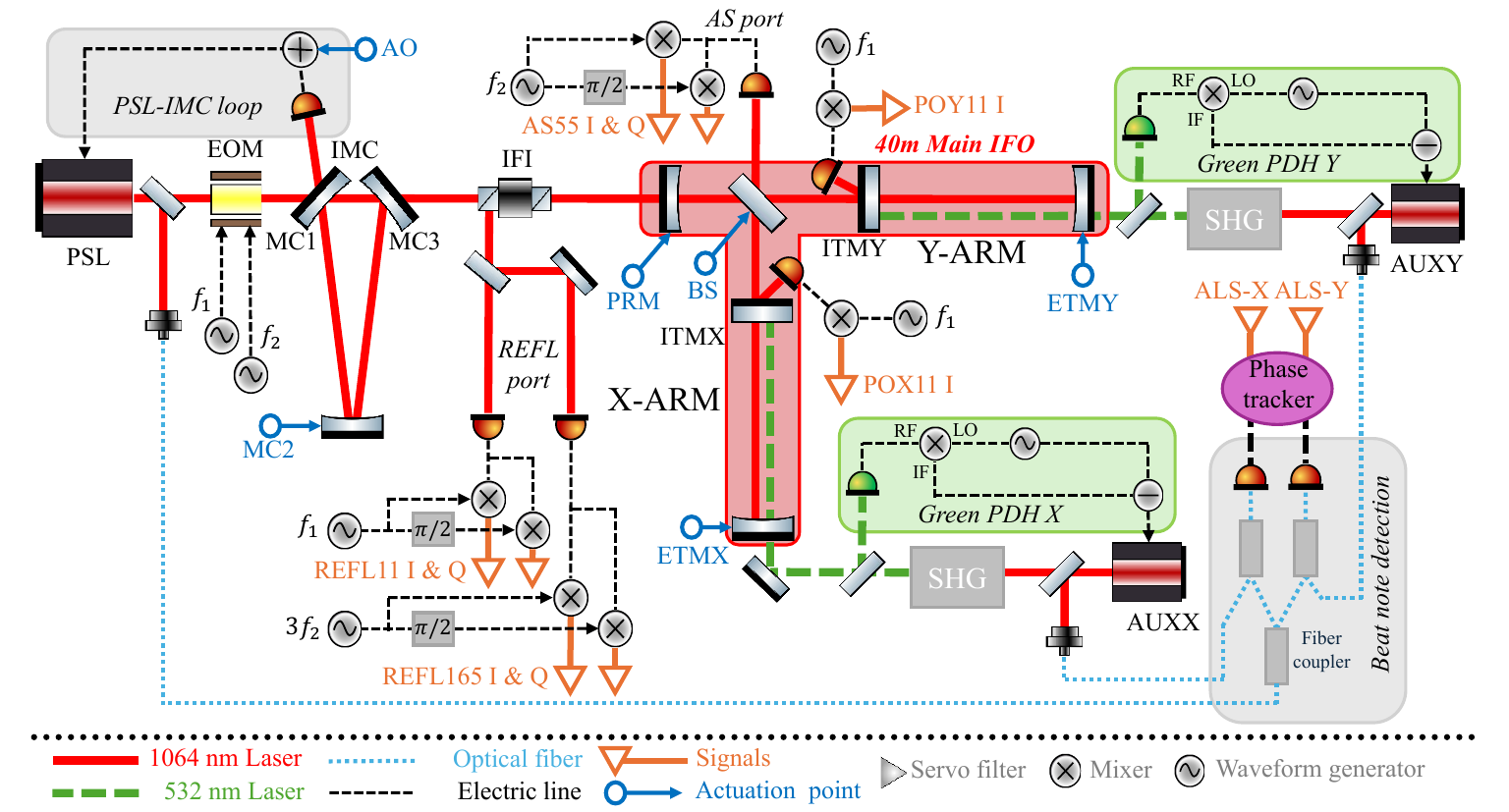}
    \caption{Optical layout of the CIT 40-meter power recycling Fabry-Perot Michelson interferometer. $f_1 = 11.066209$ MHz, and $f_2 = 5f_1$.
    ALS signals are obtained by the beat note detection between the Pre-Stabilized Laser (PSL) and the two auxiliary lasers (AUXX \& AUXY).
    These beat notes are made by picking off 1064 nm light from the three lasers and coupling them together in optical fibers.
    The two auxiliary lasers are injected into the arm cavities from the back of the end mirrors (ETMX \& ETMY), after the wavelength is halved to 532 nm by Second Harmonic Generation (SHG).
    Both AUXX and AUXY are frequency stabilized to keep them on resonance to their respective arm cavities.
    To generate the error signal from the green reflection, frequency dithering is applied to the auxiliary laser frequency.
    RF error signals are obtained at the reflection (REFL) port, extracted through the input Faraday isolator (IFI), and the Anti-Symmetric (AS) port.
    We restrict the sensors shown to only those required for lock acquisition, omitting the others.
    The CARM control is achieved by actuating the mirror MC2 - corresponding to length of the IMC.
    Since the PSL frequency is locked to the IMC length, this actuation indirectly works as the PSL frequency drive.
    In addition, the additive offset (AO) actuation provides a means to directly drive the PSL frequency, which is exploited to increase CARM control bandwidth.
    \Cref{tab:SigActuateForEachDof} lists the sensors and the actuators for 4 degrees of freedom in CIT 40-meter fast lock acquisition.
    }
    \label{fig:40mIFOlayout}
\end{figure}

In this section, we describe the lock acquisition and fast CARM offset reduction for the Power-Recycled Fabry-P\'{e}rot Michelson Interferometer (PRFPMI) configuration of the CIT 40-meter prototype interferometer. \Cref{fig:40mIFOlayout} shows a schematic of the optical layout and signal readout scheme for this prototype. 
Our fundamental change enabling fast lock acquisition is to increase the low frequency ALS CARM control gain. We achieve this by adding high bandwidth feedback to the PSL frequency actuator in parallel to the existing, lower bandwidth IMC length actuation. The resulting control loop has a unity-gain-frequency exceeding $15\,\mathrm{kHz}$.
High bandwidth ALS CARM control mitigates the contamination from CARM residual motion into the PRCL and MICH degrees-of-freedom. This permits parking CARM on its resonant position, zero offset point, while controlled by ALS without breaking 3f PRCL and MICH lock. 
Furthermore, due to the reduction in CARM offset and the suppression of noise resulting from the transition from ALS to RF operation, the frequency response of the optical plant changes. Specifically the cavity pole frequency, of the CARM coupled-cavity, reduces as power accumulates in the interferometer.
Extending the bandwidth of the CARM control loop to $15\, \mathrm{kHz}$, or higher, serves to maintain the stability of the control loop during this handover.
With high bandwidth CARM control we skip many steps in the LIGO lock acquisition procedure, primarily transitions of CARM and DARM between different error signals, leading to both time savings and a more robust CARM offset reduction procedure.

\begin{figure}
    \centering
    \includegraphics[width = 1.0\linewidth]{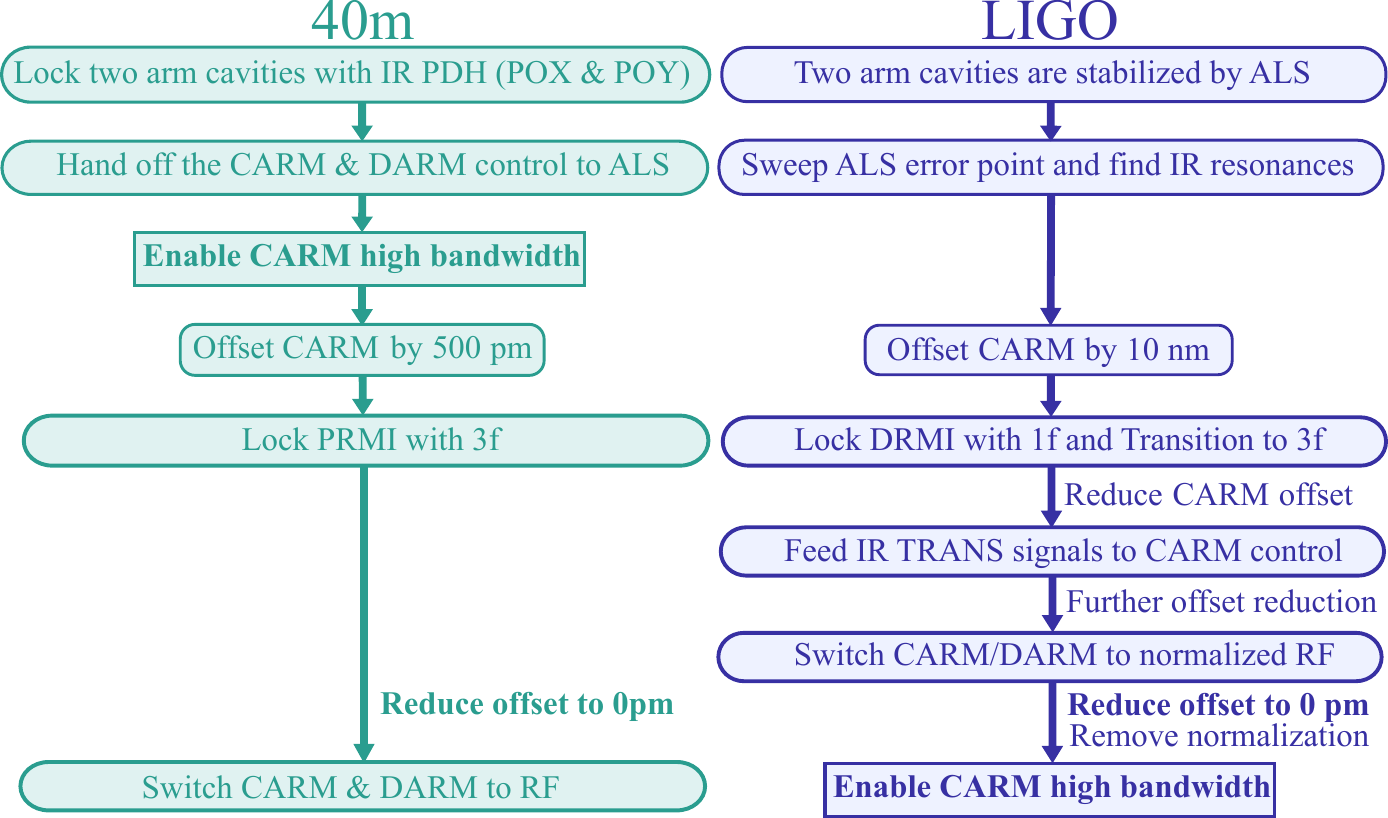}
    \caption{Flowchart for a comparison of the lock acquisition of CIT 40-meter (Left) and Advanced LIGO (Right). Main differences between the two are the following: (i) $\sim$ (iv). (i) CIT 40-meter has the high bandwidth CARM control in ALS, (ii) CIT 40-meter offsets the CARM smaller than LIGO, (iii) CIT 40-meter skips IR TRANS signals feedback during the offset reduction, (iv) LIGO has the signal recycling cavity.}
    \label{fig:FlowChart}    
\end{figure}

\begin{figure}
    \centering
    \includegraphics[width=\linewidth]{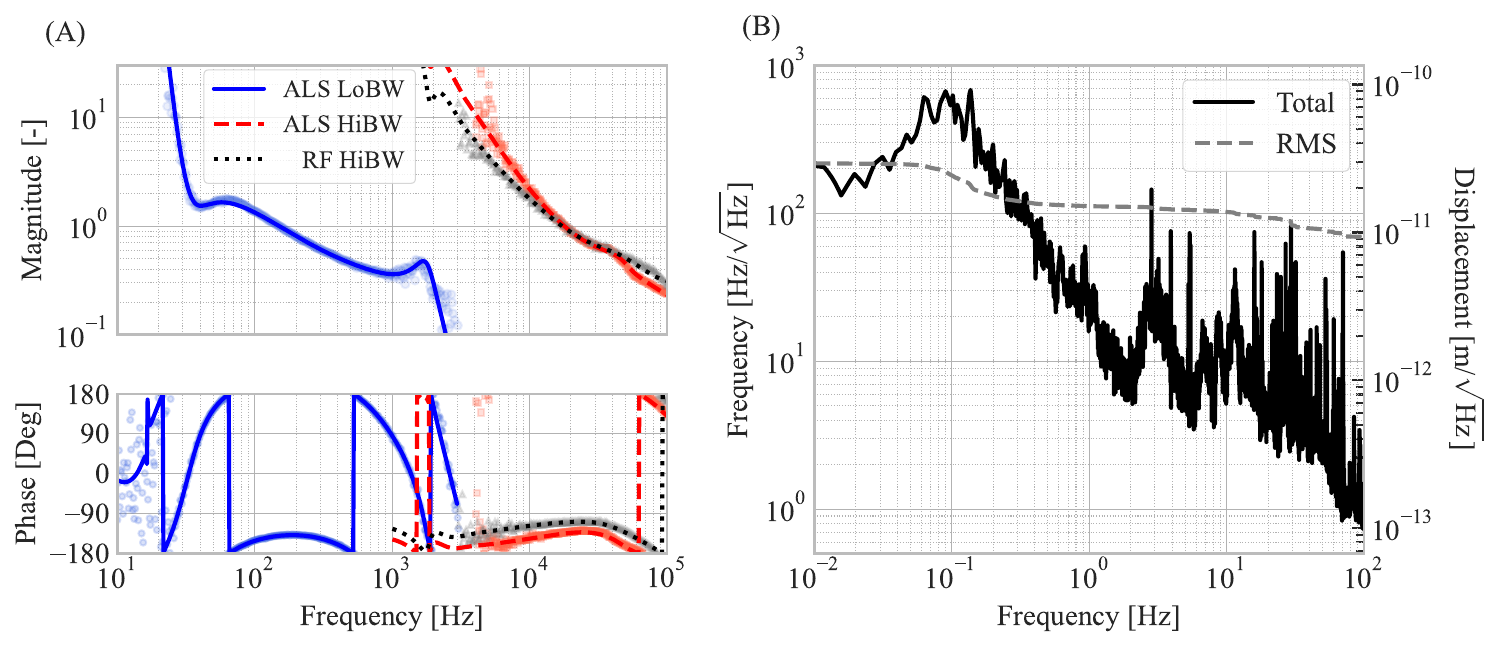}
    \caption{(A). Open loop transfer function of the ALS-CARM control and the RF-CARM control in the CIT 40-meter prototype lock acquisition. The unity gain frequency (UGF) of the ALS low bandwidth loop (Blue solid line) is limited to 150 Hz due to the slow data-processing speed of our control digital system. Enabling the additive offset path (see \Cref{fig:40mIFOlayout}), the control bandwidth is increased to 17.9 kHz in the ALS high bandwidth loop (Red dashed line). Finally, the RF high bandwidth control (Black dot line) is acquired, switching the control signal from the ALS to REFL 11 I so that the UGF is kept. A discrepancy was observed between the main loop model and the measured data outside the control bandwidth, particularly above 25 kHz. The reason for this discrepancy is not clear, but vector fitting~\cite{772353, 1645204, 4530747} successfully identified the loop pole locations, supporting the fact that they are stable. (B). Residual noise of ALS CARM control (Black) assessed by IR PDH signals as out-of-loop sensors (POX + POY). This noise does not change regardless of whether the ALS CARM control is done with the low bandwidth (Blue in the Bode plot) or the high bandwidth (Red in the Bode plot). This fact means that the error signal of the ALS CARM control is limited by sensing noises in the frequency band. Therefore, we can expect lower noise operation if the green finesse of the arm cavity is increased, as discussed in \Cref{sec:fastLIGO}.}
    \label{fig:OLTF_and_NoiseSpectrum}
\end{figure}

\begin{figure*}
    \includegraphics[width = 1.0\linewidth]{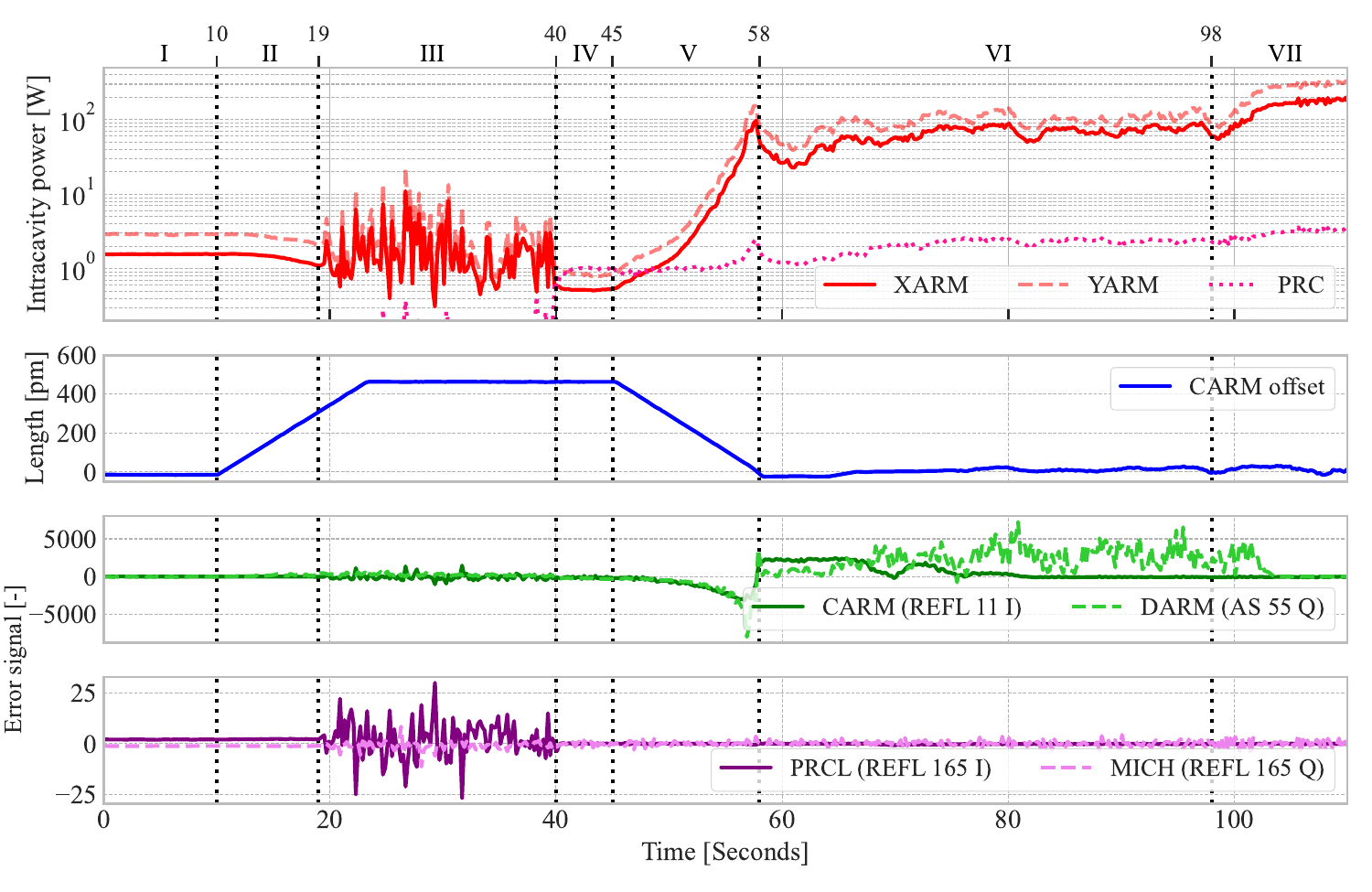}
    \caption{Transition of optical powers and error signals during PRFPMI lock acquisition in the CIT 40-meter prototype.
    The red lines show the intracavity powers of the XARM cavity, the YARM cavity, and the power recycling cavity.
    Intracavity powers of arm cavities are calibrated based on the transmitted powers of XARM and YARM, and the transmittances of ETMX and ETMY (13.7 $\pm$ 3 ppm). Also, the intracavity power of the Power-Recycling Cavity (PRC) is calibrated from the optical power transmitted by a pick-off mirror placed between the PRM and the BS, and its transmission (972 $\pm$ 59 ppm).
    The blue line shows the CARM offset in units of length, which is calibrated from the out-of-loop ALS CARM signal.
    The green and purple signals are the RF error signals for the 4 degrees of freedom in CIT 40-meter PRFPMI. REFL $f_1$ I, AS $f_2$ Q, REFL $3f_2$ I, and REFL $3f_2$ Q are used for CARM, DARM, PRCL, and MICH, respectively. 
    }
    \label{fig:40mArmPower}    
\end{figure*}

The left (CIT 40m) and right (LIGO) of \Cref{fig:FlowChart} compare the lock acquisition procedure for the CIT 40-meter prototype and aLIGO. First the CIT 40-meter begins by misaligning the power recycling mirror (PRM), and stabilizing the lengths of the two Fabry-P\'{e}rot arm cavities using IR PDH signals from POX (Pick-off from XARM) and POY (Pick-off from YARM). A digital combination of POX and POY is used to form CARM and DARM modes. Next the CARM and DARM control is handed over to ALS after matching the ALS and IR offsets at their error points. For comparison LIGO commences locking with the ALS, and scans the arm lengths to find a length where both green and IR light are co-resonant.
Both methods result in green and IR light co-resonant within the Fabry-P\'{e}rot arm cavities -- starting with IR locking is a more reliable method of reaching this co-resonant condition but produces negligible time savings.

Next the CIT 40-meter enables the additive offset (AO) path and increases the ALS CARM control bandwidth, from 150 Hz up to 25 kHz. To affect this handoff ALS CARM control is transitioned from the LIGO real time control and data system (RTCDS) \cite{BORK2021advLIGOrts} to a digitally configured FPGA system.
\Cref{fig:OLTF_and_NoiseSpectrum} (A) compares the open-loop transfer functions between these two configurations. Low bandwidth, LIGO style RTCDS, control is shown in blue, and high bandwidth, digitally configured FPGA, control is shown in red.
With CARM and DARM controlled by ALS and CARM configured in high bandwidth POX and POY become out-of-loop sensors for residual arm displacement. \Cref{fig:OLTF_and_NoiseSpectrum} (B) shows this residual motion extracted by the POX and POY sensors, both in frequency detuning and residual displacement units, with the cumulative RMS plotted as a dashed line.

After this, we add an offset to CARM to isolate MICH \& PRCL error signals from CARM length fluctuations.
This offset is currently $\sim 500\, \mathrm{pm}$, less than the CIT 40m prototype's arm cavity pole of $590\, \mathrm{pm}$.
Compared to LIGO, this offset is $20 \times$ smaller and does not completely drive the arm cavities off-resonance and is satisfactory for PRMI lock acquisition.
To enable PRC locking the PRM is brought back into alignment, and then both MICH and PRCL are locked directly with 3f-demodulated error signals \cite{arai2001robust} measured at the reflection port.
Once all four degrees-of-freedom are locked, by ALS and 3f controls, the CARM offset is ramped back to zero. Finally CARM and then DARM are switched from ALS to RF error signals.

\Cref{fig:40mArmPower} shows the time evolution of the arm cavity circulating power during PRFPMI lock acquisition, with high bandwidth ALS, CARM control (ALS-HiBW). During period I, the two arm cavities are stabilized near resonance by this ALS control configuration. In parts II and III a 500 pm CARM offset is added to the controller's error point, shifting the arm cavity lengths slightly away from IR resonance. This reduces the coupling between the CARM and DARM degrees-of-freedom, and those of PRMI (PRCL \& MICH). Throughout segment III, and while the CARM offset is still increasing to 500 pm, the PRM is aligned back and resonant flashes are observed in the Power-Recycling Cavity (PRC). During period IV, PRCL and MICH are locked using 3f-demodulated signals - their use prevents PRMI lockloss during the CARM offset reduction process. This loss of PRMI control would otherwise occur due to sign changes in the effective, compound reflectivity of the arm cavity mirrors, dependent on whether the arm cavities are on or off resonance~\cite{ARAI200015, arai2001robust}.

During period V, the CARM offset is reduced to 0 pm in a single ramp, bringing the arm cavities onto resonance and increasing the intracavity power due to the PRC. High-bandwidth ALS CARM control maintains the residual motion well within the Power Recycled CARM linewidth, no intermediate control transitions are required. This eliminates the multi-stage CARM offset reduction used in LIGO, and the associated failure modes where a mid-sequence lockloss forces a full restart.
In part VI, the CARM control is transferred from ALS to the RF signal - note the REFL 11 I signal (green solid line) zeroing in \Cref{fig:40mArmPower}. In this segment DARM remains controlled by low-bandwidth ($\sim100\, \mathrm{Hz}$) ALS, causing the intracavity powers to drift.
Finally in period VII, the DARM control is then transferred to the RF control, by ramping up and down the RF signal and the ALS signal for 5 seconds (98 $\sim$ 103 seconds) - observe the AS 55 Q signal (green dashed line) zeroing in \Cref{fig:40mArmPower}. The precise difference in the zero-crossing point between ALS and RF error signals results in an increase in intracavity power when locking transitions to RF error signals. Additionally intracavity power fluctuations are reduced in later parts of segment VII because RF error signals are more sensitive than their ALS counterparts.

In total, the combination of steps V, VI and VII, as per \Cref{fig:40mArmPower}, is the CARM offset reduction and the RF handover procedure for the CIT 40-meter prototype.
Its equivalence to the LIGO procedure, described in \Cref{subsec:LIGOLockSum}, is readily summarized by \Cref{fig:FlowChart}.
At the end of both processes, CARM and DARM are both locked with RF signals, and are ready to transition central interferometric control to 1f-demodulated signals.
The total time for the CARM offset reduction and the RF handover of the CARM and DARM controls, at the CIT 40-meter prototype, is 58 seconds from starting the CARM offset reduction (at 45 seconds) to finishing the RF handover of DARM (at 103 seconds).
This is split between 13 seconds for step V, 40 seconds for step VI, and 5 seconds for step VII.
Additional improvements could be made in steps VI and VII which each include a wait time. The first ensures that lockloss does not occur at zero CARM offset, and the subsequent confirms that the full RF CARM control has been completed, again without lockloss. Future speed-up can be achieved by reducing these pauses and shortening ramp-up and down process.

Compared with median, one-shot LIGO times in \Cref{tab:LIGOlockAcq}, and \Cref{fig:LIGOlockAcq}, 58 seconds is substantially faster than any time at LHO and competitive with the fastest times at LLO.
Our primary speed-up comes from skipping several intermediate steps of LIGO's CARM offset reduction, reducing it directly to 0 pm in a single step. One can clearly understand this by comparing \Cref{fig:40mArmPower} to Figure 5 in \cite{staley2014achieving}, which shows the intracavity power evolution in aLIGO during CARM offset reduction. In the CIT 40-meter prototype, the intracavity power increases rapidly over the 13 seconds of period V. In contrast in LIGO has several intervals during which the CARM offset and the intracavity powers remain constant. In LIGO, the lock is often lost in those intervals, when the control configuration is transitioning. The simplicity of the CIT 40-meter prototype lock acquisition will lead to a reduction in the number of such locklosses in the offset reduction, which will provide us with a further increase in observation time for gravitational-wave events.

\begin{table}
    \centering
    \caption{Signals and actuators for each degree of freedom in the CIT 40-meter prototype lock acquisition sequence.
    The sensors for each degree of freedom are listed in order, with the handoff order shown in \Cref{fig:FlowChart}.
    }
    \begin{tabular}{c|c|c}
    \toprule
        DoF & Sensor & Actuator \\
    \midrule
        \multirow{2}{*}{CARM (LoBW)} & POX + POY   & \multirow{2}{*}{MC2} \\
                                     & ALSX + ALSY &  \\
                                     \hline
        \multirow{2}{*}{CARM (HiBW)} & ALSX + ALSY & \multirow{2}{*}{MC2, AO} \\
                                     & REFL 11 I   &  \\
        \hline
        \multirow{3}{*}{DARM} & POX - POY & \multirow{3}{*}{ETMX - ETMY}  \\
                              & ALSX - ALSY &  \\
                              & AS 55 Q &  \\
        \hline
        PRCL & REFL 165 I & $\sqrt{2}$PRM - BS \\
        \hline
        MICH & REFL 165 Q & PRM \\
    \bottomrule
    \end{tabular}
    \label{tab:SigActuateForEachDof}
\end{table}

\section{Discussions towards Implementation on LIGO}
\label{sec:fastLIGO}

This section discusses the applicability of our fast lock acquisition scheme to LIGO.
Stably transitioning arm cavity control from ALS to RF requires the ALS control keep the RMS of the residual CARM noise within the linewidth of the CARM coupled-cavity. When this condition is satisfied the input optical power is confined within the coupled power recycling and arm cavities, minimizing changes to the optical plant (the pole frequency of the CARM coupled-cavity). Consequentially control system transients created by switching to more sensitive RF control are stably managed.
The major challenge is the poor signal-to-noise ratio (SNR) of ALS, which stabilizes the beatnote frequencies between the PSL and the two auxiliary lasers. Its error signal has a  linear range on the order of the laser wavelength, $\lambda \sim \mathcal{O}(10^{-6} \, [\mathrm{m}])$. In contrast, the linear range of a PDH error signal is set by the cavity linewidth, i.e., scaling as the laser wavelength divided by the finesse, $\lambda/\mathcal{F} \sim \mathcal{O}(10^{-9} \, [\mathrm{m}])$.
Consequently, ALS provides a much wider dynamic range than PDH, but suffers from a poorer SNR.

In the following, \Cref{subsec:NoiseRequirement} clarifies the requirements regarding residual noise within the ALS control system that LIGO must satisfy to achieve rapid offset reduction and stable RF handover.
\Cref{subsec:NoiseSource} identifies potential noise sources and systematically discusses strategies for their suppression.
\Cref{subsec:GreenThermalNoise} quantitatively discusses the adverse effects that increasing green finesse for the sensing noise reduction has on coating thermal noise.

\subsection{ALS residual noise requirements}
\label{subsec:NoiseRequirement}

The linewidth of the CARM-coupled cavity is given by
\begin{align}
    \Delta f_\mathrm{CARM} = \frac{\Delta f_\mathrm{ARM}}{G_\mathrm{PR}}\, , \quad \Delta L_\mathrm{CARM} = \frac{\Delta L_\mathrm{ARM}}{G_\mathrm{PR}}
\end{align}
in frequency and length, where $\Delta f_\mathrm{ARM}, \Delta L_\mathrm{ARM}$ is the linewidth of the arm cavity in frequency and length, and $G_\mathrm{PR}$ is the gain of the power recycling cavity, interchangeably denoted power recycling gain. The linewidth of the arm cavity is
\begin{align}
    \Delta f_\mathrm{ARM} = \frac{c}{4L} \frac{1}{\mathcal{F}} \, , \quad \Delta L_\mathrm{ARM} = \frac{\lambda}{4 \mathcal{F}}
\end{align}
$c$ the speed of light, $\lambda$ the laser's wavelength, $L$ the length of arm cavities, and $\mathcal{F}$ the arm cavity finesse.
Parameters and their specific values are listed in \Cref{tab:ParametersFortyLIGO} for both LIGO and 40m prototype.
The IR finesse is nearly identical for both the 40m and LIGO, $\mathcal{F} \sim 450$. However the arm cavities in LIGO are 100 times longer than those in the 40m. That makes the linewidth of the CARM-coupled cavity 100 times narrower in frequency, while length linewidth is unchanged.
This means that the frequency (or phase) noise of the ALS CARM control needs to be 100 times smaller in LIGO, while the displacement noise remains at approximately the same level as in the 40m case.

\begin{table}
    \centering
    \caption{Main parameters of Advanced LIGO and 40m prototype. The linewidth denotes the Half Width at Half Maximum (HWHM), which is called the pole frequency of the cavity. Parameters for LIGO are based on~\cite{cahillane2021controlling}. Power recycling gains of LIGO are based on O4~\cite{capote2025aligoO4}.}
    \begin{tabular}{l|ccccc}
    \toprule
        Parameter & Symbol & \textbf{LHO} & \textbf{LLO} & \textbf{40m} & Unit\\
    \midrule
        Main laser wavelength & $\lambda$ & 1064 & 1064 & 1064 & nm \\
        Arm length  & $L$ & 3995 & 3995 & 37.79 & m \\
        Finesse for 1064 nm (IR) & $\mathcal{F}$ & 416 & 421 & 451 & - \\
        Finesse for 532 nm (Green) & $\mathcal{F}'$ & 68 & 107 & 95 & - \\
        Power recycling gain & $G_\mathrm{PR}$ & 50 & 35 & 10 & - \\
        Arm linewidth (Frequency) & $\Delta f_\mathrm{ARM}$ & 45 & 44 & 4386 & Hz\\
        Arm linewidth (Length) & $\Delta L_\mathrm{ARM}$ & 640 & 631 & 590 & pm\\
        CARM linewidth (Frequency) & $\Delta f_\mathrm{CARM}$ & 0.90 & 1.27 & 439 & Hz\\
        CARM linewidth (Length) & $\Delta L_\mathrm{CARM}$ & 12.8 & 18.0 & 59.0 & pm\\
    \bottomrule
    \end{tabular}
    \label{tab:ParametersFortyLIGO}
\end{table}

\subsection{Noise sources}
\label{subsec:NoiseSource}
\begin{figure}[htbp]
    \centering
    \includegraphics[width = 0.9\linewidth]{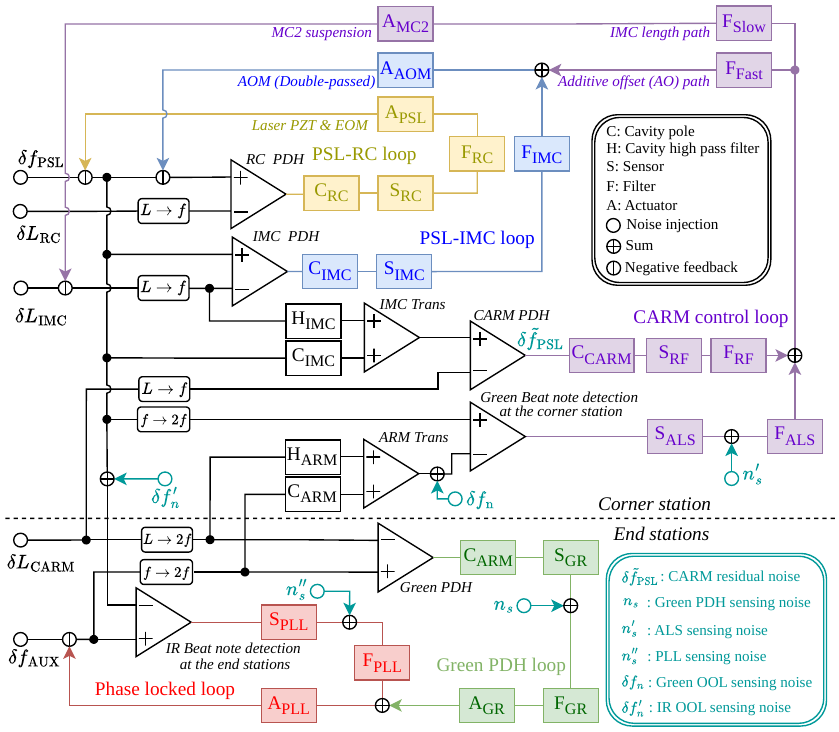}
    \caption{
    Block diagram for the CARM control in LIGO with the ALS high bandwidth control.
    This is based on the optical layouts of LIGO described in~\cite{staley2014achieving, cahillane2021controlling, Kwee:12}.
    Five loops (PSL-RC, PSL-IMC, PLL, Green PDH, CARM control) are closed for five fluctuations ($\delta f_\mathrm{PSL}, \delta f_\mathrm{AUX}, \delta L_\mathrm{RC}, \delta L_\mathrm{IMC}, \delta L_\mathrm{CARM}$).
    $\mathrm{S}, \mathrm{F}, \mathrm{A}$ are the optical gain of sensors, the servo filter, and the actuator response, respectively.
    $\mathrm{C}$ is the cavity pole due to the delay by the cavity round-trip time. $\mathrm{H}$ is the cavity high pass filter due to the Doppler shift~\cite{kiwamu2012phd}.
    Regarding the CARM control loop, two sensors, $\mathrm{S_{ALS}}$ and $\mathrm{S_{RF}}$, are used. One is the ALS signal acquired by the beat note detection and the phase tracker/the frequency discriminator.
    Another is the RF signal acquired in the REFL port.
    Then, two types of actuation, $\mathrm{A_{MC2}}$ and $\mathrm{A_{AOM}}$, are implemented for feedback control.
    The first one is the IMC length via the MC2.
    Since the PSL-IMC loop is closed with high bandwidth, this works as the frequency actuation.
    The second one is called the additive offset (AO). This enables us to increase the bandwidth more than $10 \, \mathrm{kHz}$.
    Differences between LIGO and 40m prototype are (i) to (iii). (i). 40m prototype does not have PLL, (ii). 40m prototype does not have the reference cavity. (iii). IR beat notes are used to get the ALS signal in the 40m prototype instead of green light.
    }
    \label{fig:BlockDiagram}
\end{figure}

Here, we discuss three noise sources, namely the frequency noise and displacement noise mentioned above, as well as sensing noise.
\Cref{fig:BlockDiagram} shows the block diagram of the CARM control loop if LIGO installs the ALS high bandwidth control for the fast offset reduction. In terms of the CARM control, there are three displacement noises ($\delta L_\mathrm{RC}, \delta L_\mathrm{IMC}, \delta L_\mathrm{CARM}$) and two frequency noises ($\delta f_\mathrm{PSL}, \delta f_\mathrm{AUX}$) that need to be controlled.
$\delta L_\mathrm{RC}, \delta L_\mathrm{IMC}, \delta L_\mathrm{CARM}$ are the cavity displacement noises of the reference cavity, the IMC, and the arm cavities. Note that there are two arm cavities, but that they are represented in the diagram as the CARM noise.
$\delta f_\mathrm{PSL}, \delta f_\mathrm{AUX}$ are the frequency noise of PSL and AUX lasers. Similarly, note that there are two auxiliary lasers, AUXX and AUXY, but that they are represented in the diagram as a common mode of frequency noise.
For these five noises, five control loops are engaged: the PSL-RC loop, the PSL-IMC loop, the phase locked loop (PLL), the green PDH loop, and the CARM control loop.
The PSL-RC and PSL-IMC loops work so that the main laser frequency $\delta f_\mathrm{PSL}$ tracks the length fluctuations of the reference cavity and the input mode cleaner $\delta L_\mathrm{RC}, \delta L_\mathrm{IMC}$~\cite{Kwee:12}. Note that the 40m prototype does not have the reference cavity, as described in \Cref{fig:40mIFOlayout}, and that both LIGO and 40m prototype have a pre-mode cleaner cavity (PMC), but this is abstracted here.
The PLL locks the AUX laser frequency to the PSL frequency by stabilizing their beat note at the reference oscillation frequency.
The green PDH loop drives the oscillation frequency, and ensures that the auxiliary laser frequency $\delta f_\mathrm{AUX}$ tracks the CARM fluctuations of the arm cavities $\delta L_\mathrm{CARM}$.
Due to the four feedback loops, in a frequency band where high control gains are realized, the two frequency noises, $\delta f_\mathrm{PSL}, \delta f_\mathrm{AUX}$, can be regarded as a copy of the length noise of the reference cavity \& the IMC, and the arm cavity, respectively, except for the influence of sensing noise.
The main contributions to length noise are seismic noise, radiation pressure noise, and thermal noise.
Seismic noise is determined by the performance of the isolation stacks and the suspension system, and is the dominant contribution to this length noise at low frequencies.
Radiation pressure and thermal distortions are sufficiently reduced during lock acquisition by operating the interferometer at low power, as is already being done at LIGO and other GW detectors.
The other loop, the CARM control loop, is for locking the PSL frequency, $\delta f_\mathrm{PSL}$, to the CARM fluctuation, $\delta L_\mathrm{CARM}$.
While the RF sensor directly detects the difference between the CARM motion and the PSL frequency, the ALS sensor detects the difference between the auxiliary laser frequency and the PSL frequency.
However, except for the sensing noises, the ALS signal indirectly sees the same thing as the RF signal due to the green PDH loop.
The requirement for the stable handoff is that $\delta \Tilde{f}_\mathrm{PSL}$ in \Cref{fig:BlockDiagram} is suppressed enough in advance, before starting the CARM handover to RF ($\mathrm{F_{RF}} = 0$).

Even if the five loops have high open-loop gains, the residual noise, $\delta \Tilde{f}_\mathrm{PSL}$, is limited by the sensing noises.
The sensing noises are categorized into five types as shown in \Cref{fig:BlockDiagram}.
$n_s, n_s'$, and $n_s''$ are the sensing noise of Green PDH locking, ALS beatnote detection, and PLL.
$\delta f_n$ and $\delta f_n'$ are out-of-loop (OOL) noises for the green and IR, respectively.
These noise sources originate outside the control loops and arise during the transmission of laser light between the corner station and the end stations, which are 4 km apart. $\delta \Tilde{f}_\mathrm{PSL}$ is given by the sensing noises as follows.
\begin{align}
    \label{eq:SensingNoise}
    \delta \Tilde{f}_\mathrm{PSL} \simeq\delta f_\mathrm{n} - \frac{n_s}{\mathrm{S_{GR}}} - \frac{n_s'}{\mathrm{S_{ALS}}}
\end{align}
where $\mathrm{S_{GR}}, \mathrm{S_{ALS}}$ are the sensing gains of the green PDH and the ALS.
Here, we assumed that open loop gains are significantly greater than unity for the five loops, which is true in the low frequencies where the RMS of $\delta f_\mathrm{PSL}$ is dominated.
As a result, the $\delta f_n'$ and $n_s''$ are suppressed by the Green PDH loop.
For the $\delta \Tilde{f}_\mathrm{PSL}$ given by \Cref{eq:SensingNoise}, the RMS of $\delta \Tilde{f}_\mathrm{PSL} < \Delta f_\mathrm{CARM}$ is required to confine all the laser power into the arm cavities and PRC.

For the first term in \Cref{eq:SensingNoise}, the dominant noise source in LIGO is fringe-wrapping noise, caused by slow drifts in the optical path length between the ITMs and the corner-station photodetector for the green beat note~\cite{staley2014achieving}.
The second and third terms of $\delta \Tilde{f}_\mathrm{PSL}$ in \Cref{eq:SensingNoise} can be reduced by improving sensing gains $\mathrm{S_{GR}}$ and $\mathrm{S_{ALS}}$.
First, one can improve $\mathrm{S_{ALS}}$ by increasing the powers picked off from the PSL and AUX lasers, because it is proportional to their optical powers $P_\mathrm{PSL}$ and $P_\mathrm{AUX}$.
\begin{align}
        \mathrm{S_{ALS}} \propto P_\mathrm{PSL} P_\mathrm{AUX}
\end{align}
To improve the sensing gain $\mathrm{S_{GR}}$, one could make the arm cavity critically coupled for 532 nm. This design allows all incident light to transmit through the cavity while minimizing reflected light, which leads to reducing scattered light in the green reflection path that interferes with PDH locking. Furthermore, for the same reason, it is important to verify and improve the mode matching ratio of the green laser to the arm cavity. Finally one could increase the green laser power and/or the arm cavity green finesse, because the PDH discriminant is proportional to finesse.
\begin{align}
        \mathrm{S_{GR}} \propto \mathcal{F}'
\end{align}
where $\mathcal{F}'$ is the green finesse of the arm cavity. Currently $\mathcal{F}'$ is 68 at LHO and 104 at LLO~\cite{cahillane2021controlling}. The following subsection discusses the implications of redesigning ITM/ETM coatings for lower 532 nm transmission, and thus higher $\mathcal{F}'$.

\subsection{Thermal Noise Considerations for Increasing Green Finesse}
\label{subsec:GreenThermalNoise}

Any proposed changes to test mass coatings must be designed to maintain detector performance during observation. Although increasing the green finesse improves ALS control as described before, it must be limited so as not to significantly worsen coating thermal noise (CTN) and degrade gravitational wave sensitivity.

The current aLIGO coating stacks consist of alternating layers of low-index silica ($\mathrm{SiO_2}$) and high-index titania-doped tantala ($\mathrm{Ti:Ta_2O_5}$)~\cite{PhysRevD.91.042002.2015}. The aLIGO coatings achieve sub-ppm absorption and high reflectivity for 1064 nm, but coating Brownian noise is currently the limiting noise source for the detector's mid-band sensitivity \cite{capote2025aligoO4}. At $100$ Hz, the aLIGO Brownian noise ASD is $1.08 \times 10^{-20}\ \text{m}/\sqrt{\text{Hz}}$ and the thermo-optic noise is $2.17 \times 10^{-21}\ \text{m}/\sqrt{\text{Hz}}$ (calculated from the pygwinc noise budgeting package v0.6.2 \cite{pygwinc}, using the baseline Advanced LIGO parameter configuration \cite{aasi2015advanced}). The noise budget for aLIGO O4 estimates total coating thermal noise as $1.13 \times 10^{-20}\ \text{m}/\sqrt{\text{Hz}}$ at 100 Hz. \cite{capote2025aligoO4}.

To study the impact of increasing green finesse $\mathcal{F}'$ on CTN, we run a global dielectric coating optimizer \cite{venugopalan2024global} using aLIGO stack materials ($\mathrm{SiO_2}$ and $\mathrm{Ti:Ta_2O_5}$) to maintain existing 1064 nm performance. We evaluate the increase in CTN as we increase $\mathcal{F}'$ from 100 to 10,000. Details on the optimization cost function, inputs, and result validation can be found in \appref{sec:ctn_app}.

\begin{figure}
    \centering
    \includegraphics[width = 0.9\linewidth]{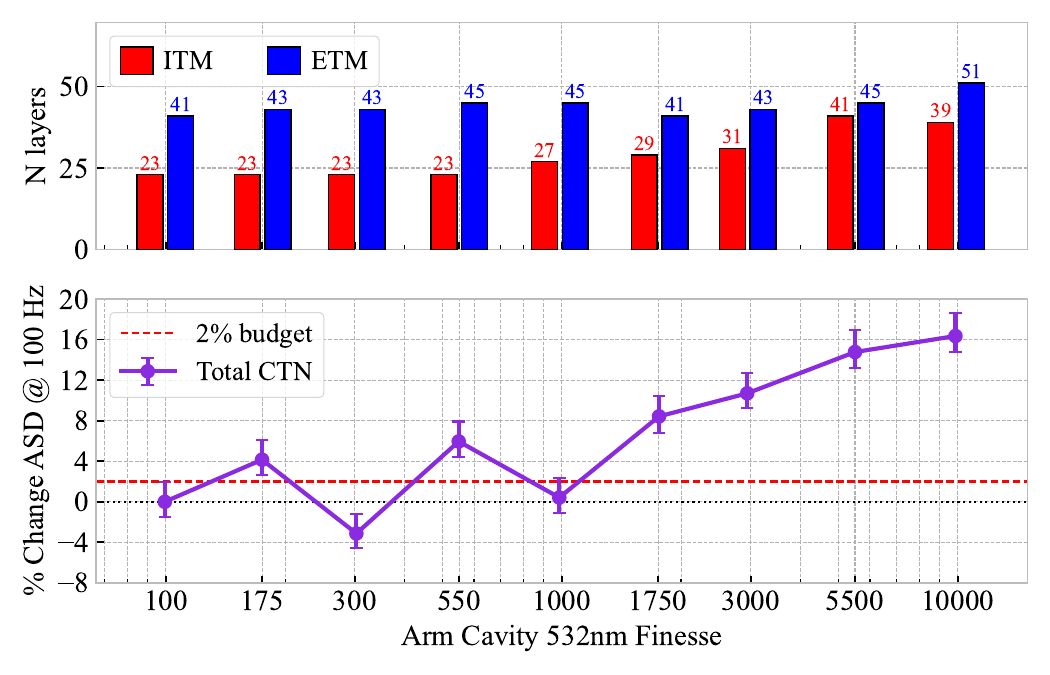}
    \caption[]{Coating thermal noise at 100 Hz vs. green finesse. Vertical error bars are derived from the MC noise posterior distributions.}
    \label{fig:total_ctn_vs_finesse}
\end{figure}

We display results for CTN vs. $\mathcal{F}'$ in \Cref{fig:total_ctn_vs_finesse}. Since the number of ITM stack layers must increase dramatically to meet lower 532 nm transmission targets, the ITM Brownian noise grows predictably as $\mathcal{F}'$ is increased. The 5 ppm ETM transmission specification for 1064 nm already sets a high baseline thickness for the ETM stack. Thus it is the ITM Brownian noise that dominates the increase in overall detector CTN with increasing $\mathcal{F}'$. We choose not to interpret the \textit{absolute} noise results from this global optimization, but rather consider the \textit{relative} increase in noise from the optimizer's result at $\mathcal{F}'=100$. We set a tolerance of 2\% on acceptable CTN increase.
From reviewing the optimization results, it is clear that there is room for increasing $\mathcal{F}'$ while keeping CTN below that 2\% tolerance. Although the optimizer returned solutions that don't fit a monotonic increase of CTN with $\mathcal{F}'$, we leave it to future dedicated coatings engineering to design low-noise, robust stack solutions. Our analysis gives confidence that the 532 nm finesse can be increased up to ~1000 while adhering to this CTN tolerance. The optimized ITM and ETM  designs for $\mathcal{F}'=1000$ are detailed in \appref{sec:ctn_app}, in \Cref{tab:coating_results} and \Cref{fig:coating_corner,fig:coating_spectral,fig:coating_noise,fig:coating_layers}.

\section{Conclusions}
\label{sec:Conclusion}

In this paper, we have presented the lock acquisition scheme demonstrated at the 40m prototype interferometer.
By controlling CARM up to the high bandwidth and achieving high loop gain in the 10-100 Hz band, we avoid that PRCL and MICH noises are contaminated through loop coupling from CARM.
Also, it provides that the stability of the CARM feedback loop is kept while the optical plant of the CARM coupled cavity changes during the offset reduction and intracavity power increase.
This allows us to omit several steps currently used in LIGO during CARM offset reduction, such as gain adjustment and feedback from the transmitted power.
As a result, the overall lock acquisition procedure is simplified and accelerated, which directly translates into an increase in the effective observing time.
There is a failure mode in which LIGO's multi-stage CARM offset reduction fails partway through, forcing a complete restart of the offset reduction sequence.
These failed attempts dominate the long tail of the lock acquisition time distribution (Yellow-shaded region in \Cref{fig:LIGOlockAcq}) and are the primary driver of lost observing time.
The main cause of this is the incompatibility between the complex steps involved in offset reduction and subtle variations in the conditions of the interferometer. such as alignment, resulting from environmental factors.
Therefore, the simple offset reduction with the CARM high bandwidth control can be expected to eliminate this failure mode.

Let us estimate how much time can be saved and how many GW events are expected to be observed if LIGO realizes the CARM high-bandwidth control and the simple offset reduction.
In LLO, based on the fact that the 40m lock acquisition time is 58 seconds, as listed in \Cref{tab:LIGOlockAcq}, the total time lost due to the long offset reduction is 11 days 14 hours 48 minutes over the whole period (O4a, O4b, and O4c), which is the accumulated time of 2148 instances of long offset reduction that happened.
With the sensitivity of LIGO O4 (a Binary Neutron Star range of $150 \sim 170 \, \mathrm{Mpcs}$), 254 non-retracted significant detection candidates are reported for 690 days of observation over O4a, O4b, and O4c~\cite{GraceDB_O4}.
In other words, an average of 0.37 candidates were reported per day, which suggests that we could detect 4.28 more candidates in O4 if the CARM offset reduction process had been more robust.
Beyond O5, should even higher sensitivity be achieved, accelerating this lock acquisition process is more important, as it will lead to the detection of more gravitational wave events.
Even outside of observing runs, this procedure will assist engineering improvements and detector characterization efforts by speeding up lock acquisition.

Towards the improvement of the observing time, in section \ref{sec:fastLIGO}, we discussed how the fast lock acquisition can be applied for LIGO.
Compared to the 40 m interferometer, LIGO has cavity linewidths that are roughly two orders of magnitude narrower, and therefore requires ALS control at significantly lower noise levels.
Possible strategies include further isolation from environmental and technical noise sources, such as seismic motion and electronic noise, as well as improving the signal-to-noise ratio of the green lock by increasing the 532 nm laser power and/or the finesse of the arm cavities at 532 nm.
In addition, placing the green laser source on the beamsplitter side and taking the beat with the PSL in free space before the arm cavity, rather than through a long optical fiber, can mitigate fiber-induced phase noise.

As a plan using the 40 m interferometer, we also highlight the implementation of Balanced Homodyne Detection (BHD).
LIGO plans to further increase the input laser power in O5 and beyond, which will enhance the contrast defect leaking to the antisymmetric (AS) port due to arm asymmetries.
BHD enables DC control and readout of DARM while coherently canceling this excess carrier field.
Moreover, by eliminating the need for DARM detuning required for conventional DC readout, BHD avoids the associated increase in quantum noise injected from the signal recycling cavity, providing further motivation for its adoption.
The current 40m aims to achieve DARM readout with BHD as well as LIGO, after obtaining stable alignment with the ASC system based on wavefront sensors and installing and locking two output mode cleaners.
Moreover, it is worth noting that LIGO features a signal recycling cavity, whereas the 40m prototype does not. To lock the DRMI, it is necessary to wait for two cavities (PRC and SRC) to reach the resonance point simultaneously, and because this is a stochastic event, it is expected to take longer than locking the PRMI. For 40m, we plan to bring in an SRM and develop a guide using Reinforcement Learning that will enable the DRMI locking time to be shortened.

\section*{Acknowledegments}
MO gratefully acknowledges the support of the stipend from the Japan Society for the Promotion of Science (JSPS) through the JSPS Research Fellowship for Young Scientists (DC1), and the support of the LIGO visitor program.
RB gratefully acknowledges support from the Roy T. Eddleman Quantum Innovation Fund.
FSC acknowledges generous support from the Barish--Weiss prize postdoctoral fellowship at Caltech.
RXA gratefully acknowledges support and guidance from Fred Blum.
This material is based upon work supported by NSF’s LIGO Laboratory, which is a major facility fully funded by the US National Science Foundation. 
LIGO was constructed by the California Institute of Technology and Massachusetts Institute of Technology with funding from the NSF and operates under NSF Cooperative Agreement PHY–2309200.
This research was supported in part by the Gordon and Betty Moore Foundation Grant No. 9221.
Also, we gratefully acknowledge technical support from Helen Schwartz, Jancarlo Sanchez, and Christopher Wipf, Mayank Chaturvedi.

\appendix
\renewcommand{\thesection}{\Alph{section}}
\titleformat{\section}
  {\bfseries}
  {Appendix~\thesection}
  {1em}
  {}

\section{Analysis of LIGO Lock Acquisition Segments}
\label{sec:LIGO_app}

Here we present specifics of the analysis outlined in \Cref{subsec:LIGOCarmOff}. This outlines our processing of the Guardian state data in more detail than the main text, where the results and implications are emphasized. 

At both LIGO sites the specific supervisory Guardian node managing the interferometers' overall state is called ISC\_LOCK \cite{LHO_ISCgrd,LLO_ISCgrd}. Despite the common naming the structure differs at each observatory. \Cref{tab:GRDstates} provides the names and state number relevant for CARM offset reduction during observing run four at each site - these remained unchanged for the duration of this observing run which was split over three sub-runs.

Guardian state numbers are recorded sixteen times per second. These records are then compressed into minimum, mean and maximum trends once per second, and also once per minute. We chose mean second trend data as a trade-off between sub-second precision, data retrieval time and data storage considerations - this choice doesn't affect the minute precision results provided in this paper. Mean, second trend data for the state number of ISC\_LOCK was downloaded for O4a, O4b and O4c using the start and stop times listed in \Cref{tab:OnTimes}. Removal by masking of the unavailable, NaN-padded data segments is the first step of our analysis.

\begin{table}
    \caption{State names and numbers associated with CARM offset reduction at LHO \cite{LHO_ISCgrd} and LLO \cite{LLO_ISCgrd} during observing run four. All of these states are taken from the sites' respective ISC\_LOCK Guardian node.}
    \centering
    \begin{tabular}{c|cc|cc}
        \toprule
         & \multicolumn{2}{c|}{\textbf{\makecell{LHO CARM offset reduction}}} & \multicolumn{2}{c}{\textbf{\makecell{LLO CARM offset reduction}}} \\
         & \makecell{\textbf{Commencement}} & \makecell{\textbf{Completion}} & \makecell{\textbf{Commencement}} & \makecell{\textbf{Completion}} \\
        \midrule
        \textbf{Name} & \makecell{PREP\_TR\_\\CARM} & RESONANCE & COMM\_SLEW & \makecell{CARM\_\\ZERO\_\\OFFSET} \\
        \makecell{\textbf{State}\\\textbf{Number}} & 200 & 410 & 560 & 800 \\
        \bottomrule
    \end{tabular}
    \label{tab:GRDstates}
\end{table}

\begin{table}
    \caption{Precise, second, boundaries for gravitational-wave observing runs O4a, O4b and O4c. Following the convention for this field times are provided as GPS seconds.}
    \centering
    \begin{tabular}{c|c|c}
        \toprule
        \textbf{Observing Run} & \makecell{\textbf{Start Time}\\(GPS seconds)} & \makecell{\textbf{End Time}\\(GPS seconds)} \\
        \midrule
        \textbf{O4a} & 1368975618 & 1389456018 \\
        \textbf{O4b} & 1396796418 & 1422118818 \\
        \textbf{O4c} & 1422118818 & 1447516818 \\
        \bottomrule
    \end{tabular}
    \label{tab:OnTimes}
\end{table}

Guardian states, within the ISC\_LOCK nodes, have been assigned monotonically increasing indices. For CARM offset commencement the Guardian state number must be greater than or equal to the smaller, site specific threshold. Conversely for CARM offset completion the Guardian state number is less than or equal to the greater, site specific threshold. This produces two logical value arrays, per site, which are cast as integer arrays for falling and rising edge detection to determine CARM offset reduction commencement and completion, respectively. Recovering the indices of these edges locates where, within an observing run, CARM offset reduction was commenced or completed.

The indices, generated by edge detection, are what needs to be paired to determine CARM offset reduction times. The use of second trend data creates a one-to-one correspondence between time (in seconds) and indices.
Our first and primary pairing method, accounting for all lockloss(es) and downtime associated with CARM offset reduction (\textit{with lockloss(es)}), pairs the smallest commencement time (index) which is later (larger than) the previous completion index but before (smaller than) the current completion index. This obtains the longest, maximum, duration for CARM offset reduction. Our alternative pairing method, neglecting any lockloss associated with CARM offset reduction (\textit{w/o lockloss(es)}), pairs the largest commencement time (index) which is later (larger than) the previous completion index but before (smaller than) the current completion index. Conversely this determines the shortest, minimum, duration for CARM offset reduction.

\Cref{tab:LIGOlockAcq} presents summary statistics for the \textit{w/o lockloss} and \textit{with lockloss(es)} data sets. We chose the median over the mean, as our central measure, due to its comparative robustness to outliers. Additionally the median (50th percentile) represents the duration at which CARM offset reduction completes half of the time. Conversely we choose both sample standard deviation and Fisher-Pearson (standard deviation normalized) skewness for their sensitivity to outliers. This combination of common statistics allows for easy quantification of how lockloss inclusion affects CARM offset reduction times at LIGO by comparing our \textit{w/o lockloss} and \textit{with lockloss(es)} data sets.
\section{Coating Stack Optimization Scheme}
\label{sec:ctn_app}

Here we present details of coating stack optimization across a range of green finesse values, $\mathcal{F}'$. The global dielectric coating optimizer \cite{venugopalan2024global} takes in the number of high/low index pairs, $N_\mathrm{pairs}$, as a user-specified parameter. Note that the optimizer includes a silica ``cap" layer and thus $N_\mathrm{layers} = 2N_\mathrm{pairs}+1$. The primary terms in the optimizer's cost function are the residual squared errors from target transmission values for 1064 nm and 532 nm ($T_{1064}$ and $T_{532}$ respectively). For all optimizations, we set targets of ETM $T_{1064} = 5\, \mathrm{ppm}$ and ITM $T_{1064} = 1.4\,\%$ to maintain the current LIGO finesse for 1064 nm. The next significant cost terms include a proxy calculation for coating Brownian noise \cite{dannenberg2009coating} at 100 Hz, and thermo-optic noise \cite{evans2008thermooptic} at 100 Hz. The coating optimizer's noise cost functions were originally designed to grow quadratically above a certain reasonable threshold, and evaluate to zero below that threshold. This functional form is appropriate for an optimizer aiming to hit certain transmission targets while keeping noise capped; however since we are interested in reporting minimum thermal noise achievable for select values of $\mathcal{F}'$, we replace the noise cost functions with residual squared errors from 0. The total cost function is evaluated as
\begin{align}
        C = \prod(1 + w_i c_i),
\end{align}
where $w_i$ and $c_i$ are weights and cost functions, respectively, of the $i^{th}$ cost term. See final weights for the cost function in \cref{tab:optimizer_weights}. Thermo-optic noise and absorption minimization is already satisfied by minimization of the other cost terms; adding non-zero weights leads the optimizer astray from $T_{1064}$ and Brownian noise targets. Lastly, the output of the optimizer is a coating stack design with an evaluated cost vector. We obtain a scalar cost figure of merit by computing the magnitude of this vector.

\begin{figure}
    \centering
    \includegraphics[width = 0.85\linewidth]{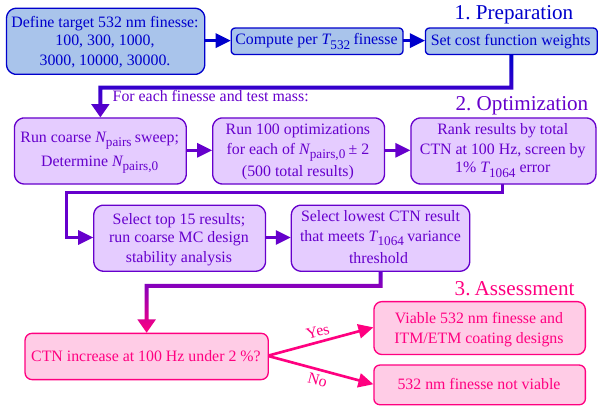}
    \caption[]{Flowchart for coating stack optimization.}
    \label{fig:ctn_flowchart}
\end{figure}

\begin{table}
    \centering
    \caption{Cost weights provided to global optimizer.}
    \begin{tabular}{l|cccccc}
    \toprule
         Cost & $T_{1064}$ & $T_{532}$ & Brownian & Thermo-optic & Absorption & E-field \\
    \midrule
        Weight & 4 & 1 & 1 & 0 & 0 & 1 \\
    \bottomrule
    \end{tabular}
    \label{tab:optimizer_weights}
\end{table}

An overview of our optimization process can be seen in \Cref{fig:ctn_flowchart}. For each value of $\mathcal{F}'$, we assume a critically-coupled cavity to determine a target $T_{532}$ for the ITM and ETM. Per $\mathcal{F}'$ and optic, we perform the following steps: we sweep over $N_\mathrm{pairs}$ and coarsely run the optimizer to find the minimum that hits a certain scalar cost threshold. This minimum is $N_\mathrm{pairs,0}.$ Next, we perform 100 refinement optimization runs for each of 5 $N_\mathrm{pairs}$ values ($N_\mathrm{pairs} = N_\mathrm{pairs,0} \pm 2$). Since only a proxy for Brownian noise was used by the optimizer, we calculate the true Brownian noise \cite{Hongetal2013} for each result at this stage. Next, the results are ranked by lowest to highest total CTN and screened by whether the resulting $T_{1064}$ is within 1 \% of target. This screening serves as a check on the mean of $T_{1064}$ (the variance is considered in the next step). The top 3 results are saved per $N_\mathrm{pairs}$, leaving up to 15 optimized stack solutions. These designs satisfy nominal requirements, but performance stability has not yet been probed. We run a coarse Monte Carlo (MC) simulation on the remaining candidates to compute the posterior spread of $T_{1064}$ and total CTN at 100 Hz, assuming a 0.5\% perturbation to high/low-n layer thicknesses and refractive indices. This analysis provides a gauge of coating design robustness to material and deposition uncertainties. We screen candidates by whether the standard deviation on $T_{1064}$, or $\sigma T_{1064}$ falls within stated aLIGO design tolerances. These specifications are $T_{1064, \mathrm{ITM}} = 1.4\% \pm 0.2\%$ \cite{harry2011coating}, and $T_{1064, \mathrm{ETM}} = 5\mathrm{ppm} \pm 1\mathrm{ppm}$ \cite{dannenberg2009coating}. Note that no ETM designs produced by the optimizer achieved $\sigma T_{1064}$ of 1 ppm --- since the MC analysis is a conservative robustness check, we accept an ETM tolerance of 1.5 ppm. We finally select the lowest CTN result that meets the stated tolerances.

\begin{table}
    \centering
    \caption{Candidate coating parameters to achieve 532 nm arm finesse of 1000. Brownian and thermo-optic terms are displayed as noise ASD at 100 Hz, in $\mathrm{m/\sqrt{Hz}}$.}
    \begin{tabular}{l|ccccccc}
    \toprule
        Optic & $N_\mathrm{layers}$ & Thickness & $T_{1064}$ & $T_{532}$ & Brownian & Thermo-optic & Absorption \\
    \midrule
        ITM & 27 & 3.9 $\mu$m & 1.41\% & 3204 ppm & 4.60e-21 & 1.26e-21 & 63.5 ppm \\
        \hline
        ETM & 45 & 7.4 $\mu$m & 5.03 ppm & 3163 ppm & 6.04e-21 & 4.93e-22 & 26.6 ppm \\
        \hline
    \bottomrule
    \end{tabular}
    \label{tab:coating_results}
\end{table}


\begin{figure}
    \centering
    \includegraphics[width=1.0\linewidth]{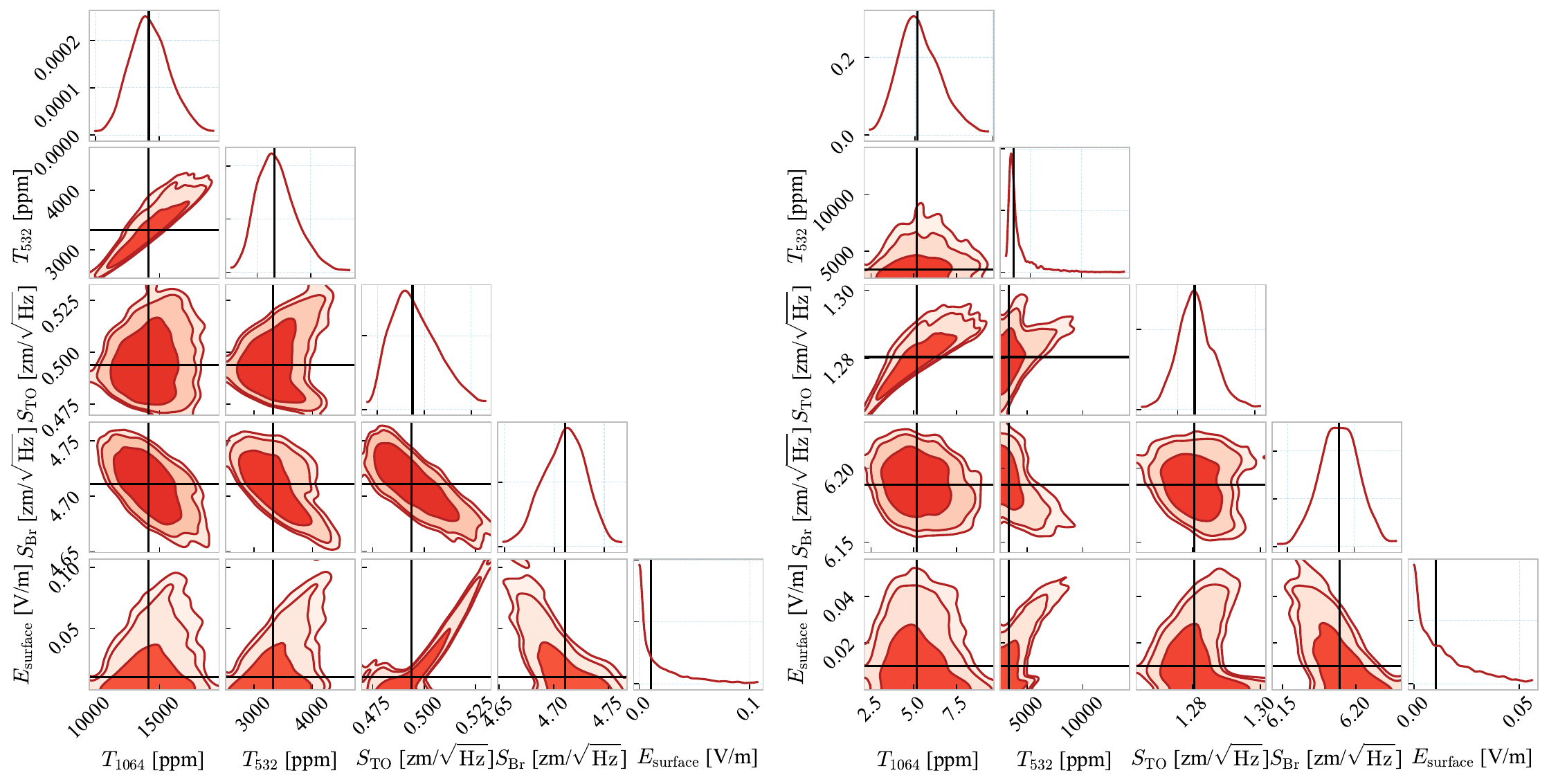}
    \caption{MCMC corner plots for robustness of ITM (left) and ETM (right) optimized stack designs. High/low-n layer thicknesses and refractive indices are perturbed by 0.5\%. Contours are shown for 68\%, 90\%, and 95\% confidence intervals. From left to right, the quantities are $T_{1064}$, $T_{532}$, thermo-optic noise, Brownian noise, and electric field at the coating surface. For conciseness, thermal noise units are written in zeptometers (zm), or $10^{-21}\mathrm{m}$.}
    \label{fig:coating_corner}
\end{figure}

\begin{figure}
    \centering
    \includegraphics[width = 1.0\linewidth]{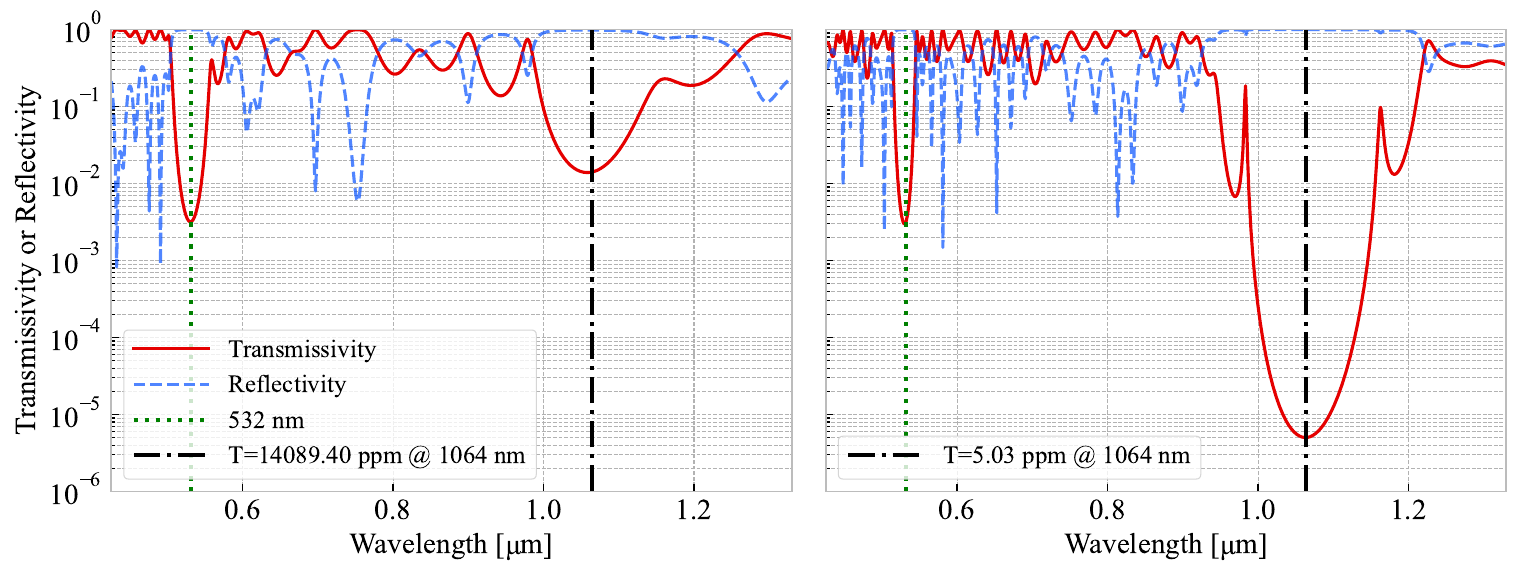}
    \caption[]{Optimized coating stack spectral response for ITM (left) and ETM (right).}
    \label{fig:coating_spectral}
\end{figure}

\begin{figure}
    \centering
    \includegraphics[width = 1.0\linewidth]{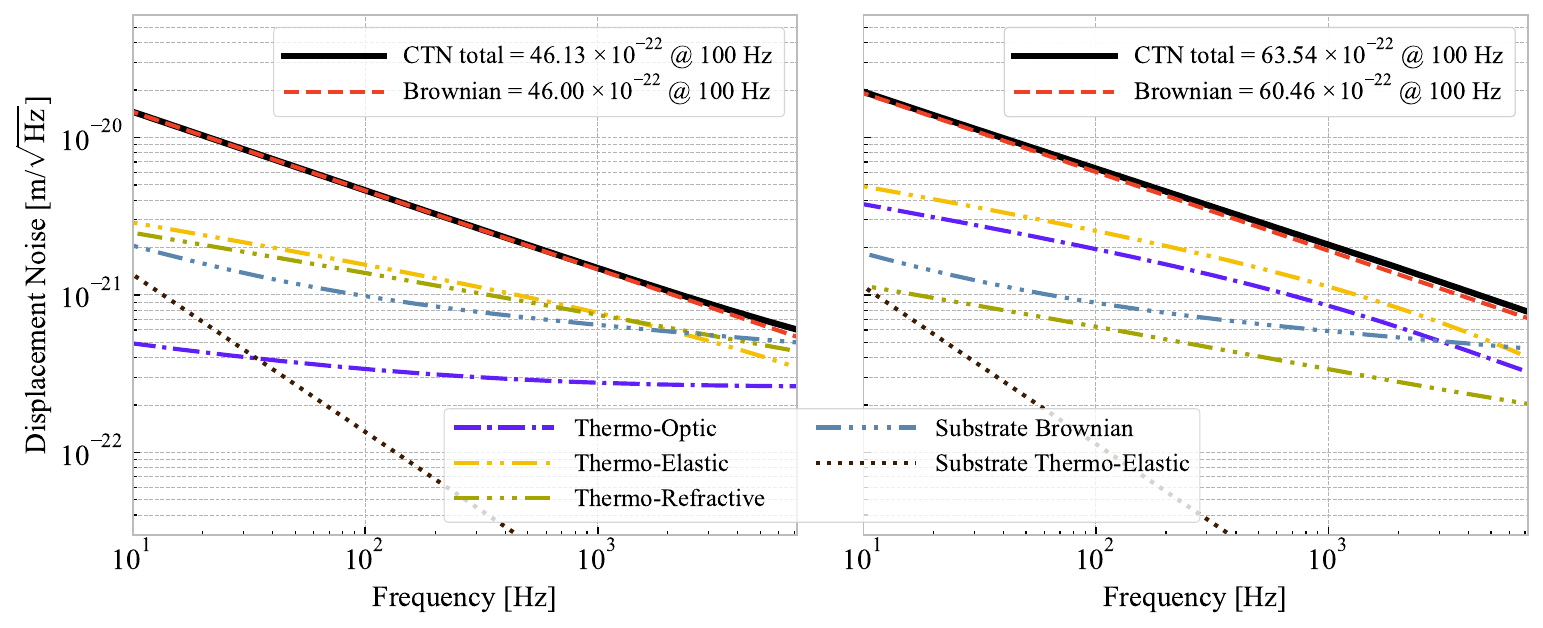}
    \caption[]{Optimized coating stack noise sources for ITM (left) and ETM (right). More information on thermal noise models in \cite{venugopalan2024global}.}
    \label{fig:coating_noise}
\end{figure}

\begin{figure}
    \centering
    \includegraphics[width = 1.0\linewidth]{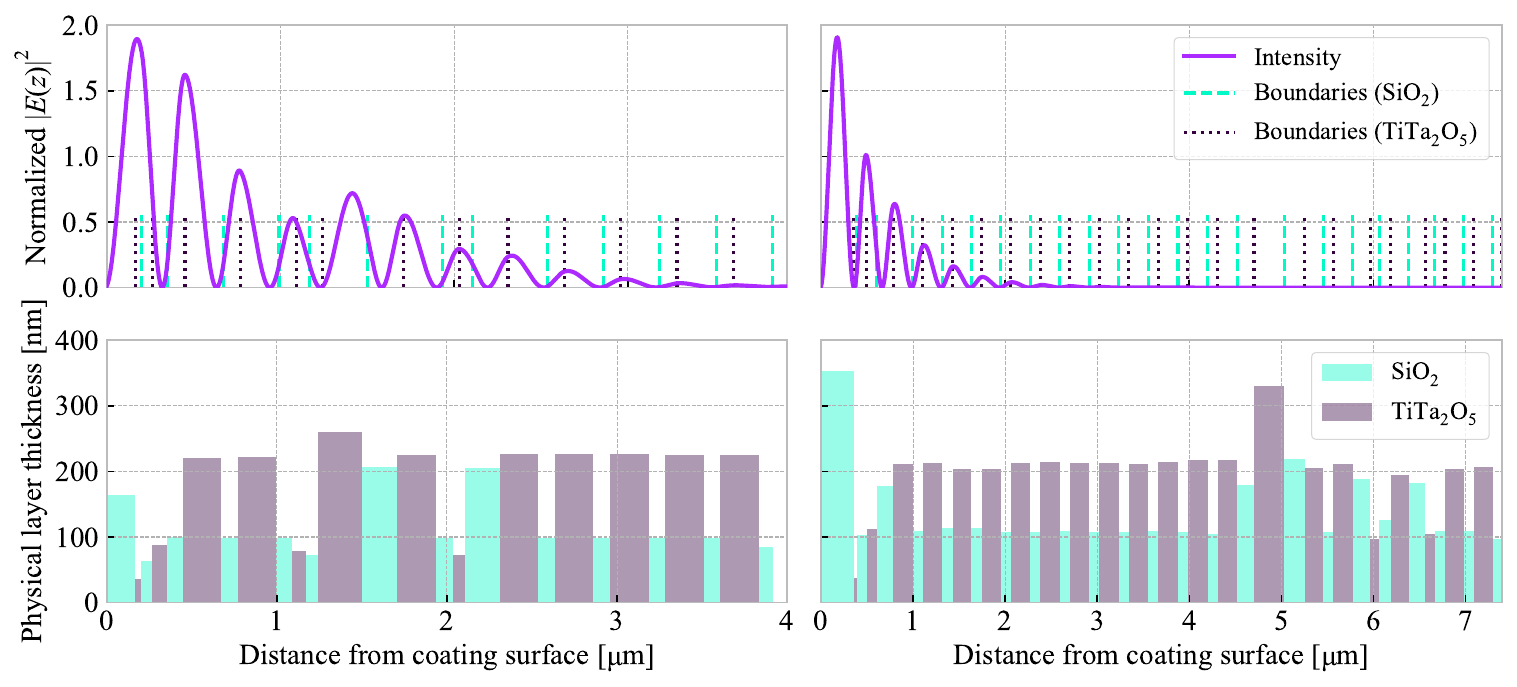}
    \caption[]{Optimized coating stack layers for ITM (left) and ETM (right). Top axes plot the normalized squared-magnitude of the electric fields inside the coating; bottom axes display the physical thickness profile of the individual dielectric layers after optimization.}
    \label{fig:coating_layers}
\end{figure}

\clearpage
\section*{References}
\bibliography{Ref_FastLockAcquisition}

@article{aasi2015advanced,
  title={Advanced {LIGO}},
  author={Aasi, Junaid and Abbott, BP and Abbott, Richard and Abbott, Thomas and Abernathy, MR and Ackley, Kendall and Adams, Carl and Adams, Thomas and Addesso, Paolo and Adhikari, RX and others},
  journal={Classical and quantum gravity},
  volume={32},
  number={7},
  pages={074001},
  year={2015},
  publisher={IOP Publishing},
  url = {https://iopscience.iop.org/article/10.1088/0264-9381/32/7/074001/meta}
}

@article{acernese2014advanced,
  title={Advanced Virgo: a second-generation interferometric gravitational wave detector},
  author={Acernese, Fet al and Agathos, M and Agatsuma, K and Aisa, D and Allemandou, N and Allocca, A and Amarni, J and Astone, P and Balestri, G and Ballardin, G and others},
  journal={Classical and Quantum Gravity},
  volume={32},
  number={2},
  pages={024001},
  year={2014},
  publisher={IOP Publishing},
  url = {https://iopscience.iop.org/article/10.1088/0264-9381/32/2/024001/meta}
}

@article{somiya2012detector,
  title={Detector configuration of {KAGRA}--the Japanese cryogenic gravitational-wave detector},
  author={Somiya, Kentaro},
  journal={Classical and Quantum Gravity},
  volume={29},
  number={12},
  pages={124007},
  year={2012},
  publisher={IOP Publishing},
  url = {https://iopscience.iop.org/article/10.1088/0264-9381/29/12/124007/meta}
}

@article{aso2013interferometer,
  title={Interferometer design of the {KAGRA} gravitational wave detector},
  author={Aso, Yoichi and Michimura, Yuta and Somiya, Kentaro and Ando, Masaki and Miyakawa, Osamu and Sekiguchi, Takanori and Tatsumi, Daisuke and Yamamoto, Hiroaki and Kagra Collaboration and others},
  journal={Physical Review D},
  volume={88},
  number={4},
  pages={043007},
  year={2013},
  publisher={APS},
  url = {https://journals.aps.org/prd/abstract/10.1103/PhysRevD.88.043007}
}

@article{abbott2017exploring,
  title={Exploring the sensitivity of next generation gravitational wave detectors},
  author={Abbott, Benjamin P and Abbott, Richard and Abbott, Thomas D and Abernathy, MR and Ackley, K and Adams, C and Addesso, P and Adhikari, Rana X and Adya, VB and Affeldt, C and others},
  journal={Classical and Quantum Gravity},
  volume={34},
  number={4},
  pages={044001},
  year={2017},
  publisher={IOP Publishing},
  url = {https://iopscience.iop.org/article/10.1088/1361-6382/aa51f4/meta}
}

@article{punturo2010einstein,
  title={The {E}instein {T}elescope: a third-generation gravitational wave observatory},
  author={Punturo, M and Abernathy, M and Acernese, F and Allen, B and Andersson, Nils and Arun, K and Barone, F and Barr, B and Barsuglia, M and Beker, M and others},
  journal={Classical and Quantum Gravity},
  volume={27},
  number={19},
  pages={194002},
  year={2010},
  publisher={IOP Publishing},
  url = {https://iopscience.iop.org/article/10.1088/0264-9381/27/19/194002/meta}
}

@article{izumi2016advanced,
  title={Advanced {LIGO}: length sensing and control in a dual recycled interferometric gravitational wave antenna},
  author={Izumi, Kiwamu and Sigg, Daniel},
  journal={Classical and Quantum Gravity},
  volume={34},
  number={1},
  pages={015001},
  year={2016},
  publisher={IOP Publishing},
  URL = {https://iopscience.iop.org/article/10.1088/0264-9381/34/1/015001}
}

@article{staley2014achieving,
  title={Achieving resonance in the Advanced {LIGO} gravitational-wave interferometer},
  author={Staley, A and Martynov, Denis and Abbott, Richard and Adhikari, RX and Arai, Koji and Ballmer, Stefan and Barsotti, Lisa and Brooks, AF and DeRosa, RT and Dwyer, S and others},
  journal={Classical and Quantum Gravity},
  volume={31},
  number={24},
  pages={245010},
  year={2014},
  publisher={IOP Publishing},
  URL = {https://iopscience.iop.org/article/10.1088/0264-9381/31/24/245010}
}

@phdthesis{martynov2015phd,
  author = {Denis V Martynov},
  school = {California Institute of Technology},
  title  = {Lock Acquisition and Sensitivity Analysis of Advanced {LIGO} Interferometers},
  year   = {2015},
  month  = {05},
  doi    = {10.7907/Z9Q81B1F},
  url    = {https://resolver.caltech.edu/CaltechTHESIS:05282015-142013480},
}

@article{capote2025aligoO4,
  title = {Advanced {LIGO} detector performance in the fourth observing run},
  author = {Capote, E. and Jia, W. and Aritomi, N. and Nakano, M. and Xu, V. and others},
  journal = {Phys. Rev. D},
  volume = {111},
  issue = {6},
  pages = {062002},
  numpages = {30},
  year = {2025},
  month = {03},
  publisher = {American Physical Society},
  doi = {10.1103/PhysRevD.111.062002},
  url = {https://link.aps.org/doi/10.1103/PhysRevD.111.062002}
}

@article{drever1983laser,
  title={Laser phase and frequency stabilization using an optical resonator},
  author={Drever, Ronald WP and Hall, John L and Kowalski, Frank V and Hough, James and Ford, GM and Munley, AJ and Ward, H},
  journal={Applied Physics B},
  volume={31},
  pages={97--105},
  year={1983},
  publisher={Springer},
  url = {https://link.springer.com/article/10.1007/bf00702605}
}

@phdthesis{masaki1998phd,
    title = {Power recycling for an interferometric gravitational wave detector},
    author = {Ando, Masaki},
    School ={The University of Tokyo},
    year = {1998},
    url = {https://granite.phys.s.u-tokyo.ac.jp/theses/ando_d.pdf}
}

@phdthesis{heinzel1999advanced,
    title = {Advanced optical techniques for laser-interferometric gravitational-wave detectors},
    author = {Heinzel, Gerhard},
    School ={Universit{\"a}t Hannover},
    year = {1999},
    url = {https://edocs.tib.eu/files/e002/265099560.pdf}
}

@article{meers1989frequency,
  title={The frequency response of interferometric gravitational wave detectors},
  author={Meers, BJ},
  journal={Physics Letters A},
  volume={142},
  number={8-9},
  pages={465--470},
  year={1989},
  publisher={Elsevier},
  URL = {https://www.sciencedirect.com/science/article/pii/037596018990515X}
}

@article{mizuno1993resonant,
  title={Resonant sideband extraction: a new configuration for interferometric gravitational wave detectors},
  author={Mizuno, Jun and Strain, Kenneth A and Nelson, Peter G and Chen, JM and Schilling, Roland and R{\"u}diger, Albrecht and Winkler, Walter and Danzmann, Karsten},
  journal={Physics Letters A},
  volume={175},
  number={5},
  pages={273--276},
  year={1993},
  publisher={Elsevier},
  URL = {https://www.sciencedirect.com/science/article/pii/037596019390620F}
}

@article{heinzel1996experimental,
  title={An experimental demonstration of resonant sideband extraction for laser-interferometric gravitational wave detectors},
  author={Heinzel, Gerhard and Mizuno, Jun and Schilling, Roland and Winkler, Walter and R{\"u}diger, Albrecht and Danzmann, Karsten},
  journal={Physics Letters A},
  volume={217},
  number={6},
  pages={305--314},
  year={1996},
  publisher={Elsevier},
  URL = {https://www.sciencedirect.com/science/article/pii/0375960196003611}
}

@phdthesis{kiwamu2012phd,
    title = {Multi-Color Interferometry for Lock Acquisition of Laser Interferometric Gravitational-wave Detectors},
    author = {Izumi,Kiwamu},
    School ={The University of Tokyo},
    year = {2012},
    url = {https://gwic.ligo.org/assets/docs/theses/izumi-thesis.pdf}
}

@article{izumi2012multicolour,
author = {Kiwamu Izumi and Koji Arai and Bryan Barr and Joseph Betzwieser and Aidan Brooks and Katrin Dahl and Suresh Doravari and Jennifer C. Driggers and W. Zach Korth and Haixing Miao and Jameson Rollins and Stephen Vass and David Yeaton-Massey and Rana X. Adhikari},
journal = {J. Opt. Soc. Am. A},
number = {10},
pages = {2092--2103},
publisher = {Optica Publishing Group},
title = {Multicolor cavity metrology},
volume = {29},
month = {Oct},
year = {2012},
url = {https://opg.optica.org/josaa/abstract.cfm?URI=josaa-29-10-2092},
doi = {10.1364/JOSAA.29.002092},
}

@article{mullavey2012als,
author = {Adam J. Mullavey and Bram J. J. Slagmolen and John Miller and Matthew Evans and Peter Fritschel and Daniel Sigg and Sam J. Waldman and Daniel A. Shaddock and David E. McClelland},
journal = {Opt. Express},
number = {1},
pages = {81--89},
publisher = {Optica Publishing Group},
title = {Arm-length stabilisation for interferometric gravitational-wave detectors using frequency-doubled auxiliary lasers},
volume = {20},
month = {Jan},
year = {2012},
url = {https://opg.optica.org/oe/abstract.cfm?URI=oe-20-1-81},
doi = {10.1364/OE.20.000081},
}

@article{rollins2016guardian,
    author = {Rollins, Jameson Graef},
    title = {Distributed state machine supervision for long-baseline gravitational-wave detectors},
    journal = {Review of Scientific Instruments},
    volume = {87},
    number = {9},
    pages = {094502},
    year = {2016},
    month = {09},
    issn = {0034-6748},
    doi = {10.1063/1.4961665},
    url = {https://doi.org/10.1063/1.4961665},
}

@article{ARAI200015,
title = {New signal extraction scheme with harmonic demodulation for power-recycled Fabry–Perot–Michelson interferometers},
journal = {Physics Letters A},
volume = {273},
number = {1},
pages = {15-24},
year = {2000},
issn = {0375-9601},
doi = {https://doi.org/10.1016/S0375-9601(00)00467-9},
url = {https://www.sciencedirect.com/science/article/pii/S0375960100004679},
author = {Koji Arai and Masaki Ando and Shigenori Moriwaki and Keita Kawabe and Kimio Tsubono},
}

@phdthesis{arai2001robust,
  title={Robust extraction of control signals for power-recycled interferometric gravitational-wave detectors},
  author={Arai, Koji},
  year={2001},
  school={The University of Tokyo},
  URL = {https://granite.phys.s.u-tokyo.ac.jp/theses/arai_d.pdf}
}

@article{arai2002tama300,
doi = {10.1088/0264-9381/19/7/383},
url = {https://doi.org/10.1088/0264-9381/19/7/383},
year = {2002},
month = {03},
volume = {19},
number = {7},
pages = {1843},
author = {K Arai and TAMA Collaboration},
title = {Sensing and controls for power-recycling of {TAMA300}},
journal = {Classical and Quantum Gravity},
}

@article{Miyakawa_2006,
doi = {10.1088/1742-6596/32/1/039},
url = {https://doi.org/10.1088/1742-6596/32/1/039},
year = {2006},
month = {mar},
volume = {32},
number = {1},
pages = {265},
author = {Osamu Miyakawa and Robert Ward and Rana Adhikari and Benjamin Abbott and Rolf Bork and Daniel Busby and Matthew Evans and Hartmut Grote and Jay Heefner and Alexander Ivanov and Seiji Kawamura and Fumiko Kawazoe and Shihori Sakata and Michael Smith and Robert Taylor and Monica Varvella and Stephen Vass and Alan Weinstein},
title = {Lock Acquisition Scheme For The Advanced LIGO Optical configuration},
journal = {Journal of Physics: Conference Series},
}

@article{Ward_2008,
doi = {10.1088/0264-9381/25/11/114030},
url = {https://doi.org/10.1088/0264-9381/25/11/114030},
year = {2008},
month = {may},
volume = {25},
number = {11},
pages = {114030},
author = {Ward, R L and Adhikari, R and Abbott, B and Abbott, R and Barron, D and Bork, R and Fricke, T and Frolov, V and Heefner, J and Ivanov, A and Miyakawa, O and McKenzie, K and Slagmolen, B and Smith, M and Taylor, R and Vass, S and Waldman, S and Weinstein, A},
title = {{DC} readout experiment at the {Caltech} 40m prototype interferometer},
journal = {Classical and Quantum Gravity},
}

@article{fricke2012dc,
  title={DC readout experiment in Enhanced LIGO},
  author={Fricke, Tobin T and Smith-Lefebvre, Nicol{\'a}s D and Abbott, Richard and Adhikari, Rana and Dooley, Katherine L and Evans, Matthew and Fritschel, Peter and Frolov, Valery V and Kawabe, Keita and Kissel, Jeffrey S and others},
  journal={Classical and Quantum Gravity},
  volume={29},
  number={6},
  pages={065005},
  year={2012},
  publisher={IOP Publishing}
}

@phdthesis{Robert2010phd,
    title = {Length Sensing and Control of a Prototype Advanced Interferometric Gravitational Wave Detector},
    author = {Ward, Robert L},
    School ={California Institute of Technology},
    year = {2010},
    url = {https://thesis.caltech.edu/5836/}
}

@phdthesis{Eric2018phd,
    title = {Improving the Performance and Sensitivity of Gravitational Wave Detectors},
    author = {Quintero, Eric Antonio},
    School ={California Institute of Technology},
    year = {2018},
    url = {https://thesis.caltech.edu/10521/}
}

@article{Akutsu_2020,
doi = {10.1088/1361-6382/ab5c95},
url = {https://doi.org/10.1088/1361-6382/ab5c95},
year = {2020},
month = {jan},
publisher = {IOP Publishing},
volume = {37},
number = {3},
pages = {035004},
author = {Akutsu, T and others},
title = {An arm length stabilization system for KAGRA and future gravitational-wave detectors},
journal = {Classical and Quantum Gravity},
}

@article{772353,
  author={Gustavsen, B. and Semlyen, A.},
  journal={IEEE Transactions on Power Delivery}, 
  title={Rational approximation of frequency domain responses by vector fitting}, 
  year={1999},
  volume={14},
  number={3},
  pages={1052-1061},
  doi={10.1109/61.772353}}

@article{1645204,
  author={Gustavsen, B.},
  journal={IEEE Transactions on Power Delivery}, 
  title={Improving the pole relocating properties of vector fitting}, 
  year={2006},
  volume={21},
  number={3},
  pages={1587-1592},
  doi={10.1109/TPWRD.2005.860281}}

@article{4530747,
  author={Deschrijver, Dirk and Mrozowski, Michal and Dhaene, Tom and De Zutter, Daniel},
  journal={IEEE Microwave and Wireless Components Letters}, 
  title={Macromodeling of Multiport Systems Using a Fast Implementation of the Vector Fitting Method}, 
  year={2008},
  volume={18},
  number={6},
  pages={383-385},
  doi={10.1109/LMWC.2008.922585}}

@Online{LLO_ISCgrd,
  author = {{LIGO Laboratory} and {LIGO Scientific Collaboration}},
  title  = {{LLO} {L1} {ISC\_LOCK.py} source code},
  url    = {https://git.ligo.org/cds/ifo/guardian/archive/l1/ISC_LOCK/-/blob/master/opt/rtcds/userapps/release/isc/l1/guardian/ISC_LOCK.py},
}

@Online{LHO_ISCgrd,
  author = {{LIGO Laboratory} and {LIGO Scientific Collaboration}},
  title  = {{LHO} {H1} {ISC\_LOCK.py} source code},
  url    = {https://git.ligo.org/cds/ifo/guardian/archive/h1/ISC_LOCK/-/blob/master/opt/rtcds/userapps/release/isc/h1/guardian/ISC_LOCK.py},
}

@techreport{dannenberg2009coating,
  author      = {Dannenberg, Rand},
  title       = {Advanced {LIGO} End Test Mass ({ETM}) Coating Specification},
  institution = {LIGO Laboratory},
  year        = {2009},
  number      = {LIGO-E0900068-v5},
  type        = {Engineering Document},
  url         = {https://dcc.ligo.org/LIGO-E0900068/public}
}

@techreport{harry2011coating,
  author      = {Harry, Gregg},
  title       = {{Advanced LIGO Input Test Mass Coating Specification}},
  institution = {LIGO Scientific Collaboration},
  year        = {2011},
  number      = {LIGO-E0900041-v6},
  url        = {https://dcc.ligo.org/LIGO-E0900041/public}
}

@techreport{evans2008thermooptic,
  author      = {Evans, Matthew},
  title       = {Thermo-Optic Noise},
  institution = {LIGO Laboratory},
  year        = {2008},
  number      = {LIGO-T080101},
  type        = {Technical Note},
  url         = {https://dcc.ligo.org/LIGO-T080101/public}
}

@article{venugopalan2024global,
  title = {Global Optimization of Multilayer Dielectric Coatings for Precision Measurements},
  author = {Venugopalan, Gautam and {Salces-C{\'a}rcoba}, Francisco and Arai, Koji and Adhikari, Rana X.},
  year = 2024,
  month = mar,
  journal = {Optics Express},
  volume = {32},
  number = {7},
  pages = {11751--11762},
  publisher = {Optica Publishing Group},
  issn = {1094-4087},
  doi = {10.1364/OE.513807},
  url = {https://opg.optica.org/oe/fulltext.cfm?uri=oe-32-7-11751},
  copyright = {\copyright{} 2024 Optica Publishing Group},
  langid = {english}
}

@article{Hongetal2013,
  author  = {Hong, Ting and Yang, Huan and Gustafson, Eric K. and Adhikari, Rana X. and Chen, Yanbei},
  title   = {Brownian thermal noise in multilayer coated mirrors},
  journal = {Physical Review D},
  year    = {2013},
  volume  = {87},
  number  = {8},
  pages   = {082001},
  doi     = {10.1103/PhysRevD.87.082001}
}

@misc{pygwinc,
  author       = {{LIGO Scientific Collaboration}},
  title        = {pygwinc: Gravitational Wave Interferometer Noise Calculator},
  year         = {2026},
  howpublished = {\url{https://git.ligo.org/gwinc/pygwinc}},
  note         = {Accessed: 2026-04-27}
}

@article{Kwee:12,
author = {P. Kwee and C. Bogan and K. Danzmann and M. Frede and H. Kim and P. King and J. P\"{o}ld and O. Puncken and R. L. Savage and F. Seifert and P. Wessels and L. Winkelmann and B. Willke},
journal = {Opt. Express},
number = {10},
pages = {10617--10634},
publisher = {Optica Publishing Group},
title = {Stabilized high-power laser system for the gravitational wave detector advanced LIGO},
volume = {20},
month = {May},
year = {2012},
url = {https://opg.optica.org/oe/abstract.cfm?URI=oe-20-10-10617},
doi = {10.1364/OE.20.010617},
}

@misc{GraceDB_O4,
    title = {{LIGO-Virgo-KAGRA Collaboration,
Gravitational-Wave Candidate Event Database}},
    url = {https://gracedb.ligo.org/superevents/public/O4/}
}

@article{acernese2008lock,
  title={Lock acquisition of the Virgo gravitational wave detector},
  author={Acernese, Fausto and Alshourbagy, M and Amico, P and Antonucci, F and Aoudia, S and Arun, KG and Astone, P and Avino, Saverio and Baggio, L and Ballardin, G and others},
  journal={Astroparticle Physics},
  volume={30},
  number={1},
  pages={29--38},
  year={2008},
  publisher={Elsevier}
}

@article{acernese2004lock,
  title={Lock acquisition of the central interferometer of the gravitational wave detector Virgo},
  author={Acernese, F and Amico, P and Arnaud, N and Babusci, D and Barille, R and Barone, F and Barsotti, L and Barsuglia, M and Beauville, F and Bizouard, MA and others},
  journal={Astroparticle Physics},
  volume={21},
  number={5},
  pages={465--477},
  year={2004},
  publisher={Elsevier}
}

@phdthesis{cahillane2021controlling,
  title={Controlling and calibrating interferometric gravitational wave detectors},
  author={Cahillane, Craig},
  year={2021},
  publisher={California Institute of Technology},
  url = {https://thesis.caltech.edu/14139/}
}

@article{BORK2021advLIGOrts,
title = {advligorts: The Advanced LIGO real-time digital control and data acquisition system},
journal = {SoftwareX},
volume = {13},
pages = {100619},
year = {2021},
issn = {2352-7110},
doi = {https://doi.org/10.1016/j.softx.2020.100619},
url = {https://www.sciencedirect.com/science/article/pii/S2352711020303320},
author = {Rolf Bork and Jonathan Hanks and David Barker and Joseph Betzwieser and Jameson Rollins and Keith Thorne and Erik {von Reis}}
}

@article{PhysRevD.91.042002.2015,
  title = {Multimaterial coatings with reduced thermal noise},
  author = {Yam, William and Gras, Slawek and Evans, Matthew},
  journal = {Phys. Rev. D},
  volume = {91},
  issue = {4},
  pages = {042002},
  numpages = {6},
  year = {2015},
  month = {Feb},
  publisher = {American Physical Society},
  doi = {10.1103/PhysRevD.91.042002},
  url = {https://link.aps.org/doi/10.1103/PhysRevD.91.042002}
}
\bibliographystyle{iopart_num.bst}

\end{document}